\documentclass[12pt]{JHEP}
\usepackage{graphicx,subfigure}
\usepackage{axodraw,pstcol,pstricks}
\usepackage{array,cite,amsfonts,amssymb,amsmath,longtable}
\usepackage{color}

\newcommand{\newc}{\newcommand}
\newcommand{\as}{\alpha_{{\rm S}}}

\newc{\gev}{\,GeV}
\newc{\mev}{\,MeV}
\newc{\ra}{\rightarrow}
\newc{\rpv}{$\mathrm{\not\!R_p}$}
\newc{\qres}{Q_{\rm res}}
\newcommand{\epem}{\rm{e}^+\rm{e}^-}

\newcommand{\CK}{{\sf CKKW}}
\newcommand{\PY}{{\sf PYTHIA}}

\newcommand{\HW}{{\sf HERWIG}}

\newcommand{\MAD}{{\sf MADGRAPH}}

\newcommand{\ee}{\rm{e}^+\rm{e}^-}
\newc{\rp}{$\mathrm{R_p}$}
\newc{\real}{\mathcal{R}e}
\newc{\alsm}{{\displaystyle \sum_{\alpha=1,2}}}
\newc{\besm}{{\displaystyle \sum_{\beta=1,2}}}
\newc{\al}{\alpha}
\newc{\sgn}{\mr{sgn}\,}
\newc{\be}{\beta}
\newc{\ga}{\gamma}
\newc{\de}{\delta}
\newc{\sla}{\!\!\!\!\!\not\:\:\!}
\newc{\slab}{\!\!\!\!\!\not\,\,\,}
\newc{\slac}{\!\!\!\!\!\!\!\not\,\,\,\,}
\newc{\met}{$\not\!\!E_T$}
\newc{\cw}{\cos\theta_W}
\newc{\sw}{\sin\theta_W}
\newc{\ssw}{\sin^2\theta_W}
\newc{\ccw}{\cos^2\theta_W}
\newc{\cbe}{\cos\beta}
\newc{\sbe}{\sin\beta}
\newc{\ort}{\frac1{\sqrt{2}}}
\newc{\sh}{\hat{s}}
\newc{\uh}{\hat{u}}
\newc{\tha}{\hat{t}}
\newc{\sa}{\sin\al}
\newc{\ca}{\cos\al}
\newc{\mz}{M_{\mr{Z}}}
\newc{\mw}{M_{\mr{W}}}
\newc{\bv}{$\mathrm{\not\!B}$}
\newc{\lv}{$\mathrm{\not\!L}$}
\newc{\beq}{\begin{equation}}
\newc{\eeq}{\end{equation}}
\newc{\ie}{{\it i.e.\/}\ }
\newc{\lam}{\lambda}
\newc{\cht}{\tilde{\chi}}
\newc{\glt}{\tilde{g}}
\newc{\upt}{\tilde{u}}
\newc{\qkt}{\tilde{q}}
\newc{\elt}{\tilde{\ell}}
\newc{\hgt}{\tilde{H}}
\newc{\nut}{\tilde{\nu}}
\newc{\dnt}{\tilde{d}}
\newc{\ftl}{\mr{\tilde{f}}}
\newc{\psb}{\bar{\psi}}
\newc{\rtt}{\sqrt{2}}
\newc{\mut}{\tilde{\mu}}
\newc{\mr}{\mathrm}
\newc{\bath}{\bar{\theta}}
\newc{\tht}{\theta}
\newc{\JC}{{\bf J}}
\newc{\lra}{\longrightarrow}
\newc{\eg}{{\it e.g.\  }}
\newc{\barr}{\begin{eqnarray}}
\newc{\earr}{\end{eqnarray}}
\newc{\me}{\mathcal{M}}
\newc{\dbm}{\partial_\mu}
\newc{\dbmu}{\stackrel{\leftrightarrow\  }{\partial^\mu}}
\newc{\sgm}{\sigma_\mu}
\newc{\captionB}[2]{\caption[{#1}]{{\small {#2}}}}
\def\Red{\red}
\renewcommand\normalcolor{\black}

\author{
 S.~Mrenna \\
Fermi National Accelerator Laboratory \\
P.O. Box 500, Batavia, IL 60510, USA {\rm and}\\
MCTP, Dept. of Physics, Univ. of Michigan \\
Ann Arbor, MI 48109, USA \\
\email{mrenna@fnal.gov}
}

\author{
 P.~Richardson \\
Theory Division, CERN \\
1211 Geneva 23, Switzerland \\
\email{Peter.Richardson@cern.ch}
}

\abstract{We report on
our exploration of matching matrix element calculations
with the parton-shower models contained in the event generators
\HW\ and \PY.  We describe results for $\epem$ collisions 
and for the hadroproduction of $W$ bosons and Drell--Yan pairs.
We compare methods based on (1) a strict implementation of ideas
proposed by Catani {\it et al.}, (2) a generalization based on
using the internal Sudakov form factors of \HW\ and \PY,
and (3) a simpler proposal of M. Mangano.  Where appropriate,
we show the dependence on various choices of scales and clustering
that do not affect the soft and collinear limits of the predictions, but
have phenomenological implications.  Finally, we comment on how to
use these results
to state systematic errors on the theoretical
predictions.
}

\keywords{QCD, Hadronic Collisions, Jets}

\title{Matching Matrix Elements \\ and Parton Showers\\
with \HW~and~\PY}

\preprint{ }

\begin{document}

\bibliographystyle{JHEP}

\section{Introduction}

  Parton-shower Monte Carlo event generators have become an 
  important tool in the
  experimental analyses of collider data. 
  These computational programs are based on the differential cross
  sections for simple
  scattering processes (usually $2\to2$ particle scatterings) together
  with a parton-shower simulation of additional QCD radiation that
  naturally connects to a model of hadronization.
  As the parton-shower algorithms are based
  on resummation of the leading soft and collinear logarithms, these
  programs may not reliably estimate 
  the radiation of hard jets, which, in turn, may bias
  experimental analyses.

  In order to improve the simulation of hard jet production in the parton
  shower, approaches were developed to correct the emission of the 
  hardest parton in an event.
  In the \PY\ event 
  generator~\cite{Sjostrand:2000wi,Sjostrand:2001yu,Sjostrand:2003wg}, 
  corrections were included for $\rm{e}^+\rm{e}^-$ 
  annihilation \cite{Sjostrand:1987hx},
  deep inelastic scattering \cite{Bengtsson:1988rw},
  heavy particle decays \cite{Norrbin:2000uu}
  and vector boson production in hadron-hadron collisions \cite{Miu:1998ju}.
  In the \HW\ event generator~\cite{Corcella:2000bw,Corcella:2002jc},
  corrections were included for $\rm{e}^+\rm{e}^-$ 
  annihilation~\cite{Seymour:1992xa},
  deep inelastic scattering \cite{Seymour:1994ti}, 
  top quark decays \cite{Corcella:1998rs} and vector boson production in 
  hadron-hadron collisions \cite{Corcella:1999gs} following the general method
  described in \cite{Seymour:1995df}.
  
  These corrections had to be calculated for each individual process and were
  only applied to relatively simple cases.
  Also, they only 
  correct the first or hardest\footnote{In PYTHIA the first emission was corrected whereas in HERWIG any emission which could be the hardest was corrected.} emission, 
  so that they give a good description of the 
  emission of one hard jet plus additional softer radiation 
  but cannot reliably 
  describe the emission of more than one
  hard jet.  Finally, they still have the same leading-order cross 
  section as the
  original Monte Carlo event generator.
Some work did address the issue of matching higher multiplicity
matrix elements and
partons showers \cite{Andre:1998vh}, but this was of limited applicability.

  Recent efforts have tried to expand upon this work.
  So far, these attempts have either provided a description of the
  hardest emission combined with a next-to-leading-order cross section
  \cite{Mrenna:1999mq,Friberg:1999fh,Collins:2000qd,Collins:2000gd,Chen:2001ci,Potter:2000an,Potter:2001ej,Dobbs:2000bx,Dobbs:2001gb,Dobbs:2001dq,Frixione:2002ik,Frixione:2002bd,Kurihara:2002ne}
   or
  described the emission of more than one hard jet~\cite{Catani:2001cc,Krauss:2002up,Lonnblad:2002sy,Lonnblad:2001iq} at leading order.

  At the same time, a number of computer programs have become  
  available~\cite{Maltoni:2002qb,Mangano:2002ea}
  which are capable of efficiently 
  generating multi-parton events in a format (the 
  Les Houches format\cite{Boos:2001cv}) that can be readily interfaced
  with \HW\ and \PY.
  In this paper, we will make use of these programs combined with the \HW\
  and \PY\ Monte Carlo event generators to implement 
  the Catani-Krauss-Kuhn-Webber~(\CK)
  algorithm suggested in  
  \cite{Catani:2001cc,Krauss:2002up} to produce a simulation of an 
  arbitrary hard process with additional hard radiation.  
  Several approaches
  are explored.  One adheres closely to the \CK\ algorithm, but
  uses \HW\ for adding additional parton showering.  The second
  is more closely tuned to the specific parton-shower generators themselves
  and calculates branching probabilities numerically (using exact
  conservation of energy and momentum) instead of
  analytically.  This is accomplished by generating pseudo-showers starting
  from the various stages of a parton-shower history.  In a later section,
  a comparison is made with a much simpler method.

As a test case, we first consider $\epem$ annihilation to jets
at $\sqrt{s}=M_Z$ using matrix element calculations with up to 6
partons in the final state.  After benchmarking this example, we
turn to the more complicated case of
hadron-hadron collisions, where we concentrate on the
production of heavy gauge bosons at the Tevatron.  
An understanding of
$W$+jet production is essential for top quark measurements
at the Tevatron, and our present tools are not adequate, with
the systematic uncertainties from event simulation rapidly 
approaching the experimental statistical uncertainties.
Improved tools will be even more important for the LHC, where
the jet multiplicities are higher.
These issues are the main motivation behind this work.

  In the next section, we present an overview of the 
parton shower--matrix element matching procedure of Catani {\it et al.}.
Section \ref{sec:ckkw} provides a detailed description of the matching
procedure
  and pays special attention to the details of
  the implementation with \HW\ and \PY. 
Section~\ref{sec:num} provides details of the pseudo--shower approach.
We then present some results for both 
  $\rm{e}^+\rm{e}^-$ and hadron-hadron collisions.
A comparison with an alternative method is presented in Sec.~\ref{sec:mlm}.
The final section contains discussion and conclusions.

Many of the later sections are devoted to the details of 
the numerical implementation of
the matching procedure, and may not be of general interest.
Those readers interested mainly in an overview and the
main results may concentrate on Sections~\ref{sec:overview}
and \ref{sec:conclusions}.

As a final note, one of the authors of the \CK\, matching algorithm 
is implementing the procedure
in the computer code
{\tt SHERPA}
\cite{Gleisberg:2003xi}.  Preliminary results have been presented for
hadron collisions well after this research, and their code is still
under development.  

\section{Overview of the Correction Procedure}
\label{sec:overview}
Parton showers are used to relate the partons produced 
in a simple, hard 
interaction characterized by a large energy scale (large means
$\gg \Lambda_{QCD}$) to the partons at an energy scale near
$\Lambda_{QCD}$.  At this lower scale, a transition is made to
a non--perturbative description of hadronic physics, with physical,
long--lived particles as the final products.  This is possible, 
because the fragmentation functions for the highly-virtual partons
obey an evolution equation that can be solved analytically or
numerically.  This solution can be cast in the form of a Sudakov
form factor, which can be suitably normalized as a probability
distribution for {\it no} parton emission between two scales.
Using the Monte Carlo method, the evolution of a parton can
be determined probabilistically, consisting of steps when the
parton's scale decreases with no emission, followed by 
a branching into sub-partons, which themselves undergo the same
evolution, but with a smaller starting point for the scale.
The evolution is ended when the energy scale of parton  reaches the
hadronization scale $\sim \Lambda_{QCD}$.  Starting from the
initial (simple) hard process, a sampling of parton showers generates
many topologies of many-parton final states, subject to certain
phase space and kinematic restrictions. 
However, the evolution equation (as commonly used)
only includes the soft and collinear fragmentation that 
is logarithmically enhanced, so that non--singular contributions
(in the limit of vanishing cut-offs) are ignored.  This means that
not enough gluons are emitted that are energetic and at a large angle from
the shower initiator, since there is no associated soft or collinear 
singularity.

In contrast, matrix element calculations give a description of a specific
parton topology, which is valid when the partons are energetic and 
well separated.  Furthermore, it includes interference between 
amplitudes with the same external partons but different internal structure.
However, for soft and collinear kinematics, the description in terms
of a fixed order of emissions is not valid, because it does not include
the interference between multiple gluon emissions which cannot be resolved.
The latter effect is known as Sudakov suppression.

The parton-shower description of hard scattering would be improved if
information from the matrix element were included when calculating
emission probabilities.  A systematic method for this can be 
developed by comparing the parton shower and matrix element predictions
for a given fixed topology.  Consider the topology of Figure 1.
The interaction consists of a hard scattering ($\epem\to \gamma/Z\to
q\bar q$) followed by a parton shower off the outgoing $\rm q\bar q$
pair.  The variables $d_i$ represent some virtuality or energy scales
that are evolved down to a cut-off $d_{\rm ini}$.  The parton
shower rate for this
given topology is a product of many factors: (1) the Born level cross
section for $\epem\to q\bar q$, (2) Sudakov form factors representing
the probability of no emission on each quark and gluon line, and
(3) the branching factors at each vertex (or splitting).
The matrix element prediction for this topology is somewhat more
complicated.  First, one needs to calculate the cross section for
the full initial- and final-state (here $\epem\to q\bar q g g q'\bar q'$).
Then, one needs to specify a particular topology.  There is no unique
way to do this, but a sensible method is to choose a clustering scheme
and construct a parton-shower history.  Ideally, the clustering variable
would be the same as the virtuality $d_i$ used to generate the parton
shower in the usual way.  Having performed the clustering, one can
then make a quantitative comparison of the two predictions.

To facilitate the comparison, we first expand the parton-shower
prediction to the same fixed order in $\alpha_s$.  This is equivalent
to setting all the Sudakov form factors to unity.  In this limit,
we see that the parton-shower product of the Born level cross section
and the vertex factors is an approximation to the exact matrix element
prediction.  As long as the values $d_i$ are all large, the
matrix element description is preferred theoretically, and the Sudakov
form factors are indeed near unity.  Therefore, the parton-shower
description can be improved by using the exact clustered matrix element
prediction.  When the values $d_i$ are {\it not} all large, and
there is a strong ordering of the value ($d_1 \gg d_2 \cdots \gg d_{\rm ini}$)
then
the parton-shower description is preferred theoretically.  
In this limit, the matrix element prediction reduces to the
product of Born level and vertex factors, provided that the argument
of $\alpha_s$ is chosen to match that used in the parton shower
(this should be related to $d_i$).  Therefore, the matrix element
prediction can be used to improve the parton-shower description
in all kinematic regions provided that: (1) the correct argument
for $\alpha_s$ is used, and (2) the Sudakov form factors are inserted
on all of the quark and gluon lines.  This provides then an {\it interpolation}
scheme between the parton shower and the matrix element prediction.
As usual, there is a systematic uncertainty associated with how
one chooses to perform the interpolation.

This provides an improvement of the specific topology considered in
Figure 1, but what of the rest of the topologies?  Matrix element
calculations can be performed for those that are simple enough,
but technically there is a limitation.  Presently, $\epem\to 6$ parton
calculations can be performed using computational farms with appropriate
cuts.  A practical solution is to choose the cut-off $d_{\rm ini}$ large
enough that the matrix element calculations in hand saturate the
dominant part of the cross section.  Then, an ordinary parton shower
can be used to evolve the parton virtualities from $d_{\rm ini}$ 
down to the hadronization scale.  It has been shown that the
correct method for doing this consists of starting the parton shower(s)
at the scale where a parton was created in the clustered topology,
and then vetoing branchings with virtualities {\it larger} than $d_{\rm ini}$
within the parton shower~\cite{Catani:2001cc}.  

The next section explains in detail the
algorithm for
implementing a
procedure like this using matrix element calculations and the
event generators \HW\ and \PY.  The results are first tested on
$\epem\to Z\to$ hadrons, and then extended to the hadronic
production of $W$ bosons.

\section{The Correction Procedure of \CK}
\label{sec:ckkw}

  We review the procedure given in \cite{Catani:2001cc,Krauss:2002up}
for correcting the parton shower of a simple process with partons
$p_a + p_b \to p_c + p_d$:

\begin{enumerate}
\setcounter{enumi}{-1}
\item Calculate (at the matrix element level) the tree level
cross sections $\sigma^0_n$ for the topology 
$p_a p_b \to p_c p_d$ $+$ $n$ additional partons for $n=0\to N$ at a resolution 
scale\footnote{The definition of the resolution parameter
  $d$ is discussed in Section~\ref{sec:jet}.} $d_{\rm{ini}}$
  using $d_{\rm{ini}}$
  as the scale for $\alpha_S$ (and the parton distribution functions
  for the case of hadronic collisions).

\item Select the jet multiplicity\footnote{In our notation $n$ will be the number of additional jets with respect to the starting process, for example $n=1$ for $\rm{e}^+\rm{e}^-\ra3\ \rm{jets}$ if the starting process is $\rm{e}^+\rm{e}^-\ra q\bar{q}$.}  $n$ with probability
\begin{equation}
   P^{(0)}_n = \frac{\sigma^0_n}{\sum^{k=N}_{k=0}\sigma^0_k}.
\end{equation}

\item Choose the distribution of the particle momenta according to the 
   matrix element
   $|\mathcal{M}_n|^2$ again using $d_{\rm{ini}}$ as the scale for 
   $\alpha_S$ (and the parton distribution functions, when appropriate).

\item Cluster the partons using the $k_T$-algorithm to determine 
   the resolution values
   $d_1>d_2\ldots>d_n>d_{\rm{ini}}$ at which $1,2,\ldots,n$ 
   additional jets are resolved.
   These give the nodal values for a tree diagram which specifies the
   $k_T$-clustering sequence for the event. In order to give a tree graph
   which could be a possible parton-shower history, only allow those
   mergings which correspond to tree-level $1\ra2$ vertices 
   in the Standard Model are allowed.
   Wherever a merging could be either a
   QCD merging or an electroweak one we take the QCD merging, 
   \eg if we merged a quark
   and an antiquark we would assume that they came from a gluon not a photon or $Z$ 
   boson.\footnote{The obvious exception to this is the last merging in $\ee$ collisions. We forbid the merging of the last $\rm q\bar{q}$ pair in $\ee$ collisions until
   all the gluons have been merged in order that we can interpret the merging history
   as a parton-shower history. Similarly in hadron collisions with outgoing
   leptons or electroweak gauge bosons the clustering of the last pair of quarks
   is forbidden until all the weakly interacting particles have been merged.}
 Examples
   of this are shown in Figs.\,\ref{fig:e+e-eg}~and~\ref{fig:W+jetseg}.

\item Apply a coupling constant reweighting 
      $\alpha_S(d_1)\alpha_S(d_2)\cdots\alpha_S(d_n)/{\alpha_S(d_{\rm{ini}})^n}\leq1$.

\item Apply a Sudakov weight factor 
  $\Delta(d_{\rm{ini}},d_j)/\Delta(d_{\rm{ini}},d_k)$
  for each internal line from a node at a scale $d_j$ to the next node
      at $d_k<d_j$. For an external line the weight factor is $\Delta(d_{\rm{ini}},d_j)$. The product of these factors gives the overall Sudakov weight.

\item Accept the configuration if the product of the $\alpha_S$ and Sudakov reweighting factors is greater than a random number $\mathcal{R}\in[0,1]$, otherwise return to step~1.

\item Generate the parton shower for this configuration, 
      vetoing all radiation with
      $d>d_{\rm{ini}}$.
      The starting scale of the shower for each parton
      is the scale at which the particle
      was created. In general, we consider the parton to have been
      created at the highest scale vertex in which it appears. There are two
      exceptions to this: in $g\ra gg$ branchings the harder gluon
      is considered to have been created at the scale of the parent gluon and the softer
      at the $g\ra gg$ vertex; the quark and antiquark produced in the
      branching $g \ra q \bar{q}$ are considered to have been produced at the
      scale of the parent gluon. Examples of how this works in practice are given in
      Figs.\ref{fig:e+e-eg}~and~\ref{fig:W+jetseg}.
\end{enumerate}

  In principle, steps 4-6 could be replaced by a reweighting of
  events when calculating
  the matrix element in step 0.  This may be more efficient, and more
  practical once a specific matching scheme is chosen.

  In general, both
  strongly and weakly interacting particles may appear in the final state.
  In applying the algorithm for these processes we take the following approach:
  in the evaluation
  of the matrix element only the strongly interacting particles will be considered in
  evaluating the resolution criterion; while in the clustering of the event to give a 
  parton-shower history all the outgoing particles will be considered. This enables us to
  work out where in the tree diagram the weakly interacting particles were produced.

  The \CK\ procedure provides a matching between the matrix element and
  parton shower at the next-to-leading-logarithm~(NLL) 
  level \cite{Catani:2001cc}.
  However there are a number
  of choices to be made which do not effect the logarithmic behaviour
  but do effect the results.
  In the rest of this section, we will discuss these choices.

\subsection{Clustering Algorithm(s): Parton-Shower History}
\label{sec:jet}
  Here, we review the $k_T$-algorithm for jet clustering
  in hadron-hadron collisions \cite{Catani:1993hr}.
  The algorithm for $\rm{e}^+\rm{e}^-$ collisions is identical except
  that no beam mergings are considered.
  The algorithm is defined in the centre-of-mass frame of the hadron-hadron
  collision, and proceeds as follows:
\begin{enumerate}

\item For every final-state particle $i$ and every pair of final-state 
      particles
      $i,j$,
      calculate  the resolution variables $d=$ $d_{iB}$ and $d_{ij}$;

\item Select the minimum value of $d$ and perform a recombination of
      the appropriate partons into a pseudoparticle;

\item Repeat the procedure from the first step for all the particles and
      pseudoparticles until all the particles and pseudoparticles have
      $d_{ij}$ and $d_{kB}$ larger than the stopping value $d_{\rm{ini}}$.
\end{enumerate}

      While there is some freedom in the definition of the
      resolution variables, it is required that they have the following 
      form in the small angle limit:
      \begin{subequations}
      \begin{eqnarray}
         d_{iB}&\simeq& E^2_i\theta^2_{iB}\simeq k^2_{\bot iB},   \\
         d_{ij}&\simeq& \min(E^2_i,E^2_j)\theta^2_{ij}\simeq k^2_{\bot ij},
      \end{eqnarray}\label{eqn:ydef}\end{subequations}  
      where $E_i$ is the energy of the particle $i$, $\theta_{iB}$ is the
      angle of the particle $i$ with respect to the beam, $k_{\bot iB}$
      is the transverse momentum of $i$ with respect to the beam, 
      $\theta_{ij}$ is the angle between the particles $i$ and $j$, and
      $k_{\bot ij}$ is the relative transverse momentum of $i$ and $j$.
      A number of possible definitions of these variables were suggested in
      \cite{Catani:1993hr} which we will discuss here.
      The first definition is:
      \begin{subequations}
      \begin{eqnarray}
        d_{iB}&=& 2E^2_i(1-\cos\theta_{iB}), \\
        d_{ij}&=& 2\min(E^2_i,E^2_j)(1-\cos\theta_{ij}).
      \end{eqnarray}\label{eqn:angle1}\end{subequations}
      This is the definition which is used in $\rm{e}^+\rm{e}^-$ collisions
      and was suggested in order for the $\rm{e}^+\rm{e}^-$ and hadron-hadron
      algorithm to be as similar as possible. 
      However, this choice is not invariant under longitudinal boosts for
      large angle emissions.  
      A longitudinal-boost invariant definition is
      \begin{subequations}
      \begin{eqnarray}
        d_{iB}&=&     p^2_{ti},\\
        d_{ij}&=&\min(p^2_{ti},p^2_{tj})R^2_{ij},
      \end{eqnarray}
      \label{eqn:angle2}
      \end{subequations}
      where $p_{ti}$ is the transverse momentum of particle $i$.
      The generalized radius is given by
      \begin{equation}
        R^2_{ij}=f(\eta_i-\eta_j,\phi_i-\phi_j),
      \end{equation}
      with $f$ being any monotonic function with the small-angle behaviour
      \begin{equation}
        f(\eta_i-\eta_j,\phi_i-\phi_j)\simeq (\eta_i-\eta_j)^2+(\phi_i-\phi_j)^2,
        \ {\rm for}\ |\eta_i-\eta_j|, |\phi_i-\phi_j|\ra 0,
      \end{equation}
      where $\eta$ is the pseudorapidity and $\phi$ the azimuthal angle.
      Suitable choices are
      \begin{equation}
        R^2_{ij}=(\eta_i-\eta_j)^2+(\phi_i-\phi_j)^2,
        \label{eqn:angle3}
      \end{equation}
       or
      \begin{equation}
        R^2_{ij}= 2\left[\cosh(\eta_i-\eta_j)-\cos(\phi_i-\phi_j)\right],
        \label{eqn:angle4}
      \end{equation}
      which additionally has the same form as occurs in the eikonal factors
      in the QCD matrix elements.

Once the minimum value of the resolution parameters is chosen, the
partons undergo recombination. If $d_{kl}$
      is the minimum, the particles $k$ and $l$ are paired to 
      form a pseudoparticle 
      according to some particular scheme,
      while if $d_{kB}$ is the minimum value the particle is included in the 
      ``beam jets.'' 
      The choice of recombination scheme is a second choice for the
      clustering prescription.
      The simplest recombination scheme, which is used in $\rm{e}^+\rm{e}^-$
      collisions, is the E-scheme, where the 
      pseudoparticle is treated as a 
      particle with momentum $\vec{p}_{ij}=\vec{p}_i+\vec{p}_j$,
      $E_{ij}=E_i+E_j$.

      A variant is the $p_t$-scheme, which
      uses the generalized radius together with a set of 
      definitions of how to calculate $p_t$, $\eta$ and $\phi$ for
      the pseudoparticle,
\begin{subequations}
\begin{eqnarray}
   p_{t(ij)} &=& p_{ti}+p_{tj},\\
\eta_{ij}   &=& \frac{p_{ti}\eta_i+p_{tj}\eta_j}{p_{tij}},\\
\phi_{ij}   &=& \frac{p_{ti}\phi_i+p_{tj}\phi_j}{p_{tij}}.
\end{eqnarray}
\end{subequations}

      The final recombination scheme considered is
      the monotonic $p_t^2$-scheme, 
      where the values of $p_{t(ij)}$ and $R^2_{(ij)k}$ 
      with the remaining particles
      are defined in terms of those for the particles $i$ and $j$ via
\begin{subequations}
\begin{eqnarray}
   p_{t(ij)}   &=& p_{ti}+p_{tj},\\
  R^2_{(ij)k} &=&  \frac{p_{ti}^2R^2_{ik}+p_{tj}^2R^2_{jk}}{p_{ti}^2+p_{tj}^2}.
\end{eqnarray}
\end{subequations}
   Further recombinations are then defined iteratively.

Note that the different choices of resolution variable and recombination schemes were developed to make contact with experimental observables.  Here, we are
interested in connecting {\it partons} to a parton-shower history in 
a quantitative way.  Since the center-of-mass energy is known in the
theoretical calculation, there is no reason to apply the requirement of
invariance under longitudinal boosts, for example.

\subsection{Sudakov Form Factors}

  An important part of the matching procedure described above 
  is the reweighting by the Sudakov form factors. 
  Here, we review some of the relevant forms of the Sudakov form factors
  found in the various Monte Carlo event generators.

\subsubsection{\HW}
  The form factors for the coherent branching process used in \HW\ are given by
\begin{equation}
\Delta^{\sf HW}_{a\ra bc}(\tilde{t}) = \exp\left\{
 -\int^{\tilde{t}}_{4t_0}\frac{dt'}{t'}\int^{1-\sqrt{\frac{t_0}{t'}}}_{\sqrt{\frac{t_0}{t'}}}
    \frac{dz}{2\pi}\alpha_S(z^2(1-z)^2t')\hat{P}_{ba}(z)
\right\},
\label{eqn:HWsud}
\end{equation}
   where $t'$ is the evolution scale (in GeV$^2$), $t_0$ is the infra-red cut-off (in
the same units), $\tilde t$ is the starting scale for
the shower, and $\hat{P}_{ba}$ are
   the unregularized DGLAP splitting functions
\begin{subequations}
\begin{eqnarray}
P_{gg} &=& C_A\left[\frac{1-z}{z}+\frac{z}{1-z}+z(1-z)  \right],\\
P_{qg} &=& T_R\left[z^2+(1-z)^2\right],\\
P_{qq} &=& C_F\frac{1+z^2}{1-z}.
\end{eqnarray}
\end{subequations}
The variable $z$ represents the fraction of energy
shared by the partons in a
$1\to 2$ branching.  The quantity $z^2(1-z)^2t'=\frac{1}{2}p_T^2$ 
represents 
one-half the square of the relative transverse momentum of the daughters with respect
to the mother's direction of motion.
  For those branchings which are divergent (in the $z$ integral),
this is exactly the form used by \HW.
  However for the branching $g\ra q\bar{q}$, which is finite, the $z$ integral in 
  Eqn.\,\ref{eqn:HWsud} is taken from 0~to~1 and the argument of $\alpha_S$ is $t'$.
  The parameter $t_0$ 
is taken in \HW\ to be
  the square of the fictitious gluon mass (which has a default value of $0.75$~GeV).
The variable $t'$ is a generalized virtuality
related to the energy of a parton $E$ and an ordering variable
$\xi$, so that $t'=E^2\xi$.  In the branching $a\to bc$,
\begin{equation} 
\xi=\frac{p_b\cdot p_c}{E_b E_c}.  
\label{eq:hw_xi}
\end{equation}
The variable $\xi$ is required to
decrease with each emission.

\subsubsection{\PY}
The \PY~Sudakov form factor (for final-state showers)
has the expression:
\begin{equation}
\Delta^{\sf PY}_{a\ra bc}(t,\tilde{t}) = \exp\left\{
 -\int_{t}^{\tilde t}\frac{dt'}{t'}\int^{\frac{1}{2}(1+\beta)}_{\frac{1}{2}(1-\beta)}
    \frac{dz}{2\pi}\alpha_S(z(1-z){t'})\hat{P}_{ba}(z)
  \theta(p_T^2-p_{T0}^2)
\right\},
\label{eqn:PYsud}
\end{equation}
where $t'$ is the virtuality of the showering parton,
$z$ is the energy fraction of a daughter with respect
to a mother, with energies defined in the c.m.s. system of the
hard scattering, and $\beta$ the velocity of the mother.
The quantity $z(1-z)t'=p_T^2$ represents the
square of the relative transverse momentum of the daughters with respect
to the mother's direction of motion in the \PY\ variables,
and the $\theta$ function requires that the minimum $p_T$ is larger
than an infrared cut-off (related to an invariant mass cut-off).
To obtain coherence effects,
the \PY~parton shower is supplemented by the requirement
that angles also decrease in the shower.
For a branching $a \to bc$ the kinematic approximation
\begin{equation}
\theta_a \approx \frac{p_{\perp b}}{E_b} + \frac{p_{\perp c}}{E_c}
\approx \sqrt{z_a (1-z_a)} m_a \left( \frac{1}{z_a E_a} +
\frac{1}{(1-z_a) E_a} \right) = \frac{1}{\sqrt{z_a(1-z_a)}}
\frac{m_a}{E_a}
\label{eqn:PYangle}
\end{equation}
is used to derive the opening angle (which is accurate at the same level
of approximation as the one in which angular ordering is derived).
This additional requirement depends on the shower history,
and it is not simple to write down an analytic expression for the
Sudakov form factor relating to a branching embedded deep within a shower.

The primary difference in the \PY\, approach is that {\it both} 
masses and angles decrease in the (time-like) shower, whereas
only the angular variable $\xi$ strictly decreases in \HW. 
Keeping the mass variable for the shower evolution is convenient
when adding matrix-element corrections.

There is one other notable difference between the \PY\ and \HW\ 
definitions of the Sudakov
form factors.  The \PY\ definition $\Delta^{\sf PY}(t,\tilde t)$ represents 
the probability of no emission between the scales $\tilde t$ and $t$.
The infrared cut-off appears as a constraint on the minimum $p_T$ of
an emission. 
The same probability is given by $\Delta^{\sf HW}(\tilde t)/\Delta^{\sf HW}(t)$
with the \HW\ definition of the Sudakov form factor. 
While it appears that the argument of $\alpha_S$ is 
different in each case, what is actually different is the evolution variable
itself. The \PY\ variable $t_{\sf PY}$ is the invariant mass squared, whereas
the \HW\ variable $t_{\sf HW}=E^2\xi$, where $E$ is the energy of the
mother parton and $\xi$ is defined in Eqn.\ref{eq:hw_xi}.  
In the soft--collinear limit, the evolution variables are 
related by $t_{\sf PY} = 2{t_{\sf HW}}{z(1-z)}$.

\subsubsection{NLL Sudakov}
 
  The NLL Sudakov form factors 
  used in \cite{Catani:2001cc} for the branching are given by
\begin{equation}
\Delta^{\sf NLL}_{a\ra bc}(t) = \exp\left\{-\int^t_{4t_0}\frac{dt'}{t'}\int^{1-\sqrt{\frac{t'}{4t}}}_{\sqrt{\frac{t'}{4t}}}\frac{dz}{2\pi}\alpha_S(t')\hat{P}_{ba}(z)\right\},
\end{equation}
  where we have made a different choice of the scale and  regularization of
  the splitting function relative  to the \HW\ form factor.
  The splitting function can be integrated to give
\begin{equation}
\Delta_{a\ra bc}(Q) = \exp\left\{-\int^Q_{Q_1=2\sqrt{t_0}}dq\Gamma_{a\ra bc}(q,Q)\right\},
\end{equation}
   with $t'\to q=\sqrt{t'}$, and the branching probabilities are given by
\begin{subequations}
\begin{eqnarray}
\Gamma_{q\ra qg}       &=& \frac{2C_F}{\pi}\frac{\alpha_S(q)}{q}
                           \left(\ln\frac{Q}{q}-\frac34\right),\\
\Gamma_{g\ra q\bar{q}} &=& \frac{N_f}{3\pi}\frac{\alpha_S(q)}{q},\\
\Gamma_{g\ra gg}        &=& \frac{2C_A}{\pi}\frac{\alpha_S(q)}{q}
                           \left(\ln\frac{Q}{q}-\frac{11}{12}\right).
\end{eqnarray}
\end{subequations}
  Here terms which vanish in the limit $q/Q\ra0$ are neglected. 
This amounts to ignoring kinematic constraints when calculating 
probabilities, which leads to the feature that 
$\Delta^{\sf NLL}$ can be greater than 1.

  There are two major sources  of difference between the NLL and \HW\ or \PY\ Sudakov
  form 
factors:
\begin{enumerate}
\item The terms which are neglected in the NLL Sudakov form factors but
      retained in \HW\ and \PY\
      ensure that the latter 
      Sudakov form factor always satisfies $\Delta\leq1$,
      whereas the NLL
      Sudakov form factors can be larger than one. When using the
      NLL Sudakov form factors we set them to one whenever $\Delta>1$.
      Alternatively, one can demand that the leading logarithm is always
      larger than the sub-leading one in the integrand (with a theta-function),
      which removes the problem.
\item The choice of scale for $\alpha_S$ 
      in all three form factors is different.
      The choice of scale in \PY\ is $p_T$, which
      is larger or equal to $k_T$, but $\alpha_S$ in
      \PY\ is evaluated at LL, not NLL.  The scale
      in \HW\ is $p_T/\sqrt{2}$, which is smaller 
      than $k_T$ in the soft limit, and $\alpha_S$ is
      evaluated at NLL. 
\end{enumerate}
   These differences both cause the \HW\ and \PY\ Sudakov form factors to be 
   smaller than the NLL ones.  This is demonstrated in Fig.\,\ref{fig:sudakov},
   which shows a comparison of the gluon and light quark Sudakov form factors
   from \HW\, and a NLL for different cut-off scales.
   The overall effect is a larger suppression
   of the higher multiplicity matrix elements for the \HW\ Sudakov form factor than
   the NLL one.

\subsection{Choice of Scales}

The \CK\ algorithm specifies certain scales for the Sudakov form factor
and $\alpha_S$.  However, in principal, the functional form of the scale
and the prefactor are not unique, and
  we have investigated a 
  number of choices.  
  One scale choice is the nodal values of $d$ (or equivalently
  the values of $k_T$ from the clustering), which should
  work well when an angular variable is used for the parton-shower
  evolution, as in \HW.
  However, we are not limited to this particular variable.
  The construction of a parton-shower history using the $k_T$-algorithm
  and a particular recombination scheme provides 
  a series of particles and pseudoparticles.
  Kinematic quantities can be constructed from the particle momenta.
  Thus, a second possible choice is
  the dot product of the four-momenta of the particles clustered, which is
  the same as the virtuality for massless particles and is 
  the choice of initial conditions for the shower normally used in \HW.
  A third choice is the virtuality of the clustered pairs, which is
  the starting point for parton showers in \PY.
  
  The chosen scale is to be used as the starting point for the
  vetoed shower in the event generator.  When reweighting by the
  Sudakov form factors, however, we allow for the possibility  
  of a prefactor to the scale, which does not affect logarithmic
  behavior, but may have a quantitative impact nonetheless.
  If we consider, for example, 
   $\ee \ra q\bar{q}$ then the choice of the starting
  scale for the \HW\ shower is $p_{q}\cdot p_{\bar{q}}$ which corresponds to 
  a scale of $\frac12k^2_T$ in terms of the $k_T$-measure. The same applies for the
  initial-state shower in Drell-Yan production at hadron colliders. 
  However, the phase space of the \HW\ shower is such that
  no emission can occur with $k^2_T$ above $\frac12\tilde{d}$,
  where $\tilde{d}$ is the scale
  variable in the parton shower. 
  Therefore, in order to produce an emission up to the scale $\tilde d$,
  the shower scale should be at least $2\tilde d$.
  We therefore leave the prefactor of the scale in the Sudakov form factors as
  a free parameter.
   Furthermore, we also allow for a minimum value of this parameter in 
   terms of the cut-off used in the shower {\tt SCLCUT},
   where {\tt SCLCUT} is the value of the matching scale~$d_{\rm{ini}}$.
   The choices are summarized below:
\begin{subequations}
\begin{eqnarray}
  d & = & {\tt QFACT(1)} \left\{\begin{array}{ccc}
                                k^2_T         & \ \ \ \ \ \ \ \ \ \ & {\tt ISCALE=1},\\
                                2p_i\cdot p_j & \ \ \ \ \ \ \ \ \ \ &{\tt ISCALE=2},\\
                                Q^2=(p_i+p_j)^2           & \ \ \ \ \ \ \ \ \ \ &{\tt ISCALE=3},
                              \end{array}\right. 
\\
d_{cut}&=&{\tt QFACT(2)}*{\tt SCLCUT},
\end{eqnarray}
\end{subequations}
   where $k^2_T$ is the $k_T$-measure for the merging, $i$ and $j$ are the particles
   which are merged and $Q$ is the virtuality of the pseudoparticle produced in 
   the merging. 
   We also allow for a minimum starting scale for the 
   final-state parton shower, and a
   minimum starting scale of the initial-state parton shower 
      \begin{subequations}
\begin{eqnarray}
d^{FSR}_{min}&={\tt QFACT(3)}*{\tt SCLCUT}, \\ 
d^{ISR}_{min}&={\tt QFACT(4)}*{\tt SCLCUT}.
\end{eqnarray}
      \end{subequations}

   Finally, we need to specify 
   the scale to be used in $\as$. The obvious choice is
   to use the same scale as in the form factors. However in \HW\ this is not
   done and the scale in $\as$ is always lower than that in the shower. Therefore
   we have left the scale for $\as$ as an additional free parameter 
and allow for a minimum value, such that
\begin{subequations}
\begin{eqnarray}
  d_{\as} & = & {\tt AFACT(1)} \left\{\begin{array}{ccc}
                                k^2_T         & \ \ \ \ \ \ \ \ \ \ & {\tt ISCALE=1},\\
                                2p_i\cdot p_j & \ \ \ \ \ \ \ \ \ \ &{\tt ISCALE=2},\\
                                Q^2           & \ \ \ \ \ \ \ \ \ \ &{\tt ISCALE=3},
                              \end{array}\right.  \\
   d^{min}_{\as} &=& {\tt AFACT(2)}*{\tt SCLCUT}.
\end{eqnarray}
\end{subequations}

The various choices of 
{\tt ISCALE}, {\tt QFACT(1-4)} and {\tt AFACT(1-2)} allow for
flexibility in matching \HW\ or \PY\ to a $k_T$-ordered shower.
It should be noted, however, that product of Sudakov form factors
and factors of $\alpha_s$ can significantly change the normalization
of the final event sample.  In practice,
all of the distributions shown in later sections
are renormalized for comparison with the standard event generators.
The variation of scales and prefactors may also affect the
truncation of the matrix element calculation, so that uncalculated
contributions are relatively more important.

\subsection{Treatment of the Highest Multiplicity Matrix Element}

The \CK\ algorithm applies the same
  procedure
  to all the matrix elements.
However, in the numerical results presented by \CK\, there is some additional 
{\it ad hoc} reweighting applied to increase the contributions from higher
multiplicities.  
The necessity for this arises because of the practical limitation
of calculating a matrix element of arbitrary multiplicity.
  The Sudakov reweighting of the matrix element 
  and the vetoed parton showers are
  performed so as not to double-count contributions
  from higher multiplicities.
  However for the highest multiplicity matrix element, 
  this is not the case.  We consider three options, denoted by
{\tt IFINAL}.
  For {\tt IFINAL=1},
  we apply the $\alpha_S$ reweighting but not 
  the Sudakov reweighting and 
  allow the parton shower to radiate freely from the
  scale at which the partons are produced. 
However, this allows the parton
  shower to produce higher $k_T$-emissions than the matrix element. 
A better choice ({\tt IFINAL=3})
  is to apply the Sudakov weights for {\it only} the internal lines, start
  the parton shower
  at the scale at which the particle was produced and -- instead of vetoing emission
  above the matching scale -- veto emission above the scale of the particle's
  last branching in the matrix element. 
Another choice ({\tt IFINAL=2}), which is slightly
  easier to implement, 
  is to only apply the Sudakov weight for the internal lines
  but start the shower at the normal scale for the parton shower 
  and apply no veto at all.

\section{Pseudo--Shower Procedure}
\label{sec:num}

The \CK\ matching algorithm envisions a vetoed parton shower
using a $k_T$-ordered parton-shower generator.  Both
\HW\ and \PY\ are {\it not} of this type.  However, both of these
generators have well--tested models of hadronization that are
intimately connected to the parton shower, and we do not wish
to discard them out of hand.  For this reason, some aspects of
the \CK\ algorithm may not be suitable to a practical application
of parton shower--matrix element matching.

\subsection{Clustering}

The naive approach to achieving a $k_T$-veto in the \HW\ or
\PY\ shower
would be to apply an internal veto on this quantity within the
parton shower itself.
To understand how this would work in practice, 
we will first review the
kinematics of the \PY\ shower for final-state radiation.  
A given branching is specified
by the virtuality of the mother $q^2$ (selected probabilistically
from the Sudakov form factor) and the energy fraction
carried away by a daughter $z$.  In terms of these quantities,
the combination $p_T^2=z(1-z)q^2$ represents the $p_T$ of the
daughter parton to the mother in the small-angle approximation.
A cut-off
$p_{T}^{\rm min}$ is determined by the minimal allowed
invariant mass $\sim$ 1 GeV.
Using Eqn.~(\ref{eqn:PYangle}),
the
requirement of decreasing angle
$\theta_b < \theta_a$ in the branching sequence
$a\to bc, b\to de$ can be reduced to
\begin{equation}
\frac{z_b (1-z_b)}{m_b^2} > \frac{1-z_a}{z_a m_a^2} ~.
\label{eq:ango}
\end{equation}
On the other hand, the $k_T$-cluster variable expressed in
the shower variables is:
\begin{subequations}
\begin{eqnarray}
 k_T^2 &  =   &  2\min(E_i,E_j)^2(1-\cos\theta_{ij}) \\ 
       &  =   &   \min\left(\frac{z}{1-z},\frac{1-z}{z}\right)m_b^2 \label{eq:kt2b} 
\end{eqnarray}
\end{subequations}

The quantity Eq.~(\ref{eq:kt2b}) would seem to be the natural
variable to use for the veto within the shower.  However, this is
not as straightforward as it may seem.
While the showering probability is determined assuming massless
daughters, the final products conserve energy and momentum.
In the soft-collinear limit, the minimum ``$k_T$'' values will
equal those obtained from $k_T$-clustering of the final-state partons,
but, in general, this will not be true.  Since we are correcting
the matrix element predictions to the soft-collinear regime of the
parton shower, this approximation is valid.  On the other hand,
the restrictions from angular ordering via Eq.~(\ref{eq:ango})
favor $z\to \frac{1}{2}$ whereas large $z$ values are more likely
to be vetoed.  
After including the fact the invariant masses are decreasing, one
can show that the first kinematically
allowed branching has $z=\frac{1}{2}$ and $m^2_{b}=m_a Q_{\rm res}$,
where $Q_{\rm res}$ is the minimum allowed $k_T$ value.
The result is a suppression in the radiation in $k_T$-cluster
distributions in the vicinity of $\qres$.  Therefore, $k_T$-clustering
may be not be the preferred clustering algorithm, and 
other clustering schemes could
be employed (see Ref.\cite{Moretti:1998qx} for an
extensive review of clustering schemes and their
applicability) that is better suited to a particular 
event generator.  
In fact, both programs uses {\it relative $p_T$} as a variable
in $\alpha_S$.
An alternative
kinematic variable closely related to the relative $p_T$ is the {\tt LUCLUS} 
measure
\cite{Sjostrand:1983am}.
According to the {\tt LUCLUS} algorithm, clustering between two partons $i$ and $j$
is given by
\begin{eqnarray}
        d_{ij}&=& 2\left( \frac{E_iE_j}{E_i+E_j}\right)^2(1-\cos\theta_{ij})
\end{eqnarray}
instead of the relation of Eqn.\,\ref{eqn:angle1}.  
Expressed in terms of the parton-shower variables,
where $z$ is the energy fraction carried by the daughter, and
$q^2$ is the squared invariant mass of the mother, $d_{ij}$
takes the form $z(1-z)q^2$ for final-state showers and $(1-z)q^2$
for initial-state showers.  In the pseudo-shower method, partons
are clustered using the {\tt LUCLUS} measure, and the internal veto
of the parton shower is performed on the parton-shower approximation
to $d_{ij}$.

\subsection{Sudakov Form Factors}

The original \CK\ procedure uses
the analytic form of the NLL Sudakov form factor.  This is
problematic for several reasons.  First, as mentioned before,
the \PY~shower is LL, which is related to the $\alpha_S$ used
in reweighting.  Second, the ``exact'' NLL analytic expression is
derived ignoring terms of order $q/Q$.
In particular,
energy and momentum are not conserved.  This explains how
the analytic Sudakov can have a value larger than 1 -- without
phase space restrictions, the subleading logarithms will become
larger than the leading logarithms.  It would be more consistent
to require that the subleading terms are less than or equal to
the leading one, but this will affect only rather large
steps in virtuality.

In applying the \CK\ algorithm to \HW, some {\it ad hoc}
tuning of scale variables
and prefactors is necessary to improve the matching.  
This is true also for an implementation using \PY.
In fact, the situation is
further complicated by the dual requirements of decreasing mass
and angle in \PY, which is not commensurate with the
analytical Sudakov form factor.  
An alternative approach is to
use the parton shower of the generator {\it itself} to calculate the
effect of the Sudakov form factors used to reweight the
matrix element prediction.  A method similar to the one described
below was used in \cite{Lonnblad:2001iq}, but it is generalized here.
It amounts to performing a
parton shower on a given set of partons, clustering the 
partons at the end of the shower, and weighting the event
by 0 or 1 depending on whether a given emission is above
or below a given cut-off.  When many partons are present
at the matrix element level, several showers may be
needed to calculate the full Sudakov reweighting.
The algorithm for constructing the Sudakov reweighting is
as follows (using $\epem\to q\bar q n$ partons as an example):

\begin{enumerate}
\item Cluster the $n+2$ partons using some scheme.
This generates a series of $n$ clustering values $\tilde d_i$ 
($\tilde d_1>\tilde d_2, \cdots$) as well
as a complete history of the shower.  Set $\tilde d_0=\infty$,
$\tilde d_{n+1}=d_{\rm ini}$ and $k=n$.  
\item Apply a parton shower to the set of $k+2$ partons, vetoing
any emissions with $d>\tilde d_{k}$.  Cluster the final-state
partons, and reweight the event by 0 if $d_{k+1}>\tilde d_{k}$,
otherwise continue.  If the weight is 0 at any
time, then stop the algorithm and proceed to the next event.
\item Use the parton-shower history to replace the two partons
resolved at the scale $\tilde d_k$ with their mother.
Rescale the event to conserve energy-momentum.
This leaves a $k-1$ parton event.  Set $k=k-1$.
If $k\ge 0 $, go to step 2.
\end{enumerate}

Equivalently, one could perform this procedure many
times for each event and reweight by a factor equal to 
the number of events that
complete the algorithm divided by the number of tries.  

To see how the algorithm works in practice, consider a 
3 parton event($\epem\to q\bar q g$).  Application of
the clustering algorithm will associate $g$ with $q$ or
$\bar q$ at the scale $\tilde d_1$.  For concreteness,
assume the ($qg$)-combination has the smallest cluster value.  
A parton shower is 
applied to the partons starting at the scale where
each parton is created (the $Z$ scale for the $q$ and $\bar q$,
and a lower scale for $g$), vetoing internally any emission
with a cluster value $d>\tilde d_{1}$.  The final-state
partons are then clustered, and the event is retained 
only if $d_2<\tilde d_2=d_{\rm ini}$.  If the event passes this test,
the set of final-state partons is saved, and
the ($qg$) pair is replaced by the mother $q$, leaving
a $q\bar q$ event.  A parton shower is applied to these partons
vetoing internally any emission with a cluster value
$d>\tilde d_0=\infty$ (i.e., no veto).  The final-state partons
are clustered, and the event is retained only if $d_1<\tilde d_1$.
If the event also passes this test, then the original set of
showered partons have been suitably reweighted.
The series of parton showers accounts for the Sudakov form factors
on all of the parton lines between the
scales $\tilde d_k$ and $\tilde d_{k+1}$, eventually forming the full Sudakov
reweighting.

\subsection{Choice of Scales}

The starting scales for
showering the individual partons should match those scales used
in the parton-shower generator.  For \PY, this choice is
the invariant mass of the parton pair $(p_i+p_j)^2\simeq 2p_i\cdot p_j$.  
For \HW, it is $p_i \cdot p_j$.
Similarly, 
the scale and order for reweighting in $\alpha_s$ should match
the generators.  \PY\ uses
the relative $p_T$ of the branching as the argument, whereas
\HW\ uses the argument $p_T/\sqrt{2}$.  
Clustering in the variable $p_T$ is convenient, because then
the nodal values from the clustering algorithm can be used directly.

\subsection{Treatment of Highest Multiplicity Matrix Element}

To treat the showering of the partons associated with the highest
multiplicity $n$ at the matrix element level, we modify one of the steps
in applying the Sudakov form factors numerically.  Namely, in the
first test for Sudakov suppression, we do not require $d_{n+1}<d_{\rm ini}$.
We loosen our requirement to be that $d_{n+1}<\tilde d_n$, so that any
additional radiation can be {\it as hard as} a radiation in the
``hard'' matrix-element calculation, but not harder.  Furthermore, we
veto emissions with $d>\tilde d_n$, instead of $d>\tilde d_{n+1}=d_{\rm ini}$.
All other steps in the algorithm are unchanged.

\section{Matching Results}

  In this section, we present matching results using
matrix element calculations from \MAD\ and parton showers from
\PY\ and \HW.
Here, we will limit ourselves to presenting some simple 
distributions to demonstrate that the results are
sensible and justify our recommended values of certain parameters and options.

The first subsection is devoted to $\epem$ collisions, where there
is no complication from initial-state radiation of QCD partons.
Also, there is a fixed center-of-mass energy, which allows a clear
illustration of the matching.  The second subsection is devoted to
the hadronic production of weak gauge bosons, with all the 
ensuing complications.

\subsection{$\rm{e}^+\rm{e}^-$ Collisions}
 
  We will first present results for $\rm{e}^+\rm{e}^-$ collisions
  at $\sqrt{s}=M_Z$. This enables
  us to study the effects of various parameters and choices while only having
  final-state radiation.
  In particular it enables to see which are the best choices for a number of parameters
  related to the scales in the Sudakov form factors and $\as$. 

  The matrix element events were generated using \MAD~\cite{Maltoni:2002qb}
  and {\sf KTCLUS} to implement the $k_T$-algorithm with the definition given
  in Eqn.\,\ref{eqn:angle1} for the $k^2_T$-measure and the E-scheme.
  Matrix element calculations of up to 6 partons are employed,
  restricted to only QCD branchings (save for the primary
  $Z\to q\bar q$ one) and only containing light flavors of quarks and
  gluons.

Our results are cast in the form of
  differential distributions with respect to $y_n$
  where $y_n=d_n/\sqrt{s}$ and $d_n$ is the value of the $k^2_T$-measure
  where the event changes from being an $n$ to an $n-1$ jet event.
  Since \HW\ and \PY\ with matrix element corrections
  give good agreement with the LEP data for these
  observables, we compare the
  results of the different matching prescriptions to output of
  these programs.

  For clarity, we recall the meaning of jet clustering
and jet resolution.  Experimentally, jet clustering is used to relate the high multiplicity
of particles observed in collisions to the theoretical objects that can be
treated in perturbation theory.  Namely, the hadrons are related to the partons
that fragment into them.  Theoretically, we can apply clustering
also to the partons with low virtuality to relate them to a higher
energy scale and to higher virtuality partons.  
In the parton-shower picture, the daughters are related to mothers by the clustering.

Jet resolution is defined in terms of a resolution variable.  It is sometimes more convenient
to make the variable dimensionless, and this can be easily achieved in $\epem$ collisions
by dividing the $k_T$ of a particular clustering by the center-of-mass energy $\sqrt{s}$
of the collisions.  The
resolution variable sets a degree of fuzziness -- structure below that energy scale is
not to be discerned.  In $\epem\to Z\to $ partons, the choice of $k_T^{res}>M_Z$ 
corresponds to only one cluster (the $Z$ itself), which is trivial and is ignored.
The first interesting result occurs when there are three or more partons present,
in which case there is a $k_T^{res}$ which separates a 2-cluster designation from
a 3-cluster designation.  This is the largest value of $k_T$ than can be
constructed by clustering all of the partons.
This particular value of $k_T^{res}$ would be denoted by
the variable $y_3 = (k_T^{res})^2/s$.  If there are only 3 partons present, there is
no choice of $k_T^{res}$ that can yield a 4-cluster designation.  However, if 4 or
more partons are present, then there is a choice of $k_T^{res}$ that is the
boundary between a 3-cluster and a 4-cluster designation.  This is the second largest
$k_T$ that is constructed in clustering, and it would be denoted by a
non-vanishing value of $y_4$.  Similarly, if 5 or more partons are present, a value
of $y_5$ can be constructed, so on and so forth.
 
\subsubsection{\HW-\CK\ Results}
Here, we show results based on applying $k_T$-clustering,
using the NLL Sudakov for reweighting, and \HW\ for performing
the vetoed parton shower.  We will demonstrate the dependence
of our results on the choices of scale variables and prefactors
before settling on an optimized choice.

  The factors {\tt QFACT(1)} and {\tt AFACT(1)} modify the scale used
in the Sudakov form factor and argument of $\alpha_S$ respectively.
{\tt QFACT(2)} sets the minimal scale in the Sudakov.
For simplicity, we set 
{\tt QFACT(1)=QFACT(2)} and {\tt AFACT(1)=AFACT(2)}.
 We also set {\tt ISCALE=1}, so that $k_T^2$ is the evolution
 variable.
  The effect of varying {\tt QFACT(1)} and {\tt AFACT(1)}
  on the parton-level differential distribution 
  $\frac1{\sigma}\frac{d\sigma}{d\log y_3}$
  with NLL Sudakov form factors is shown in Fig.\,\ref{fig:nllscales}.
  The standard \HW\, prediction with the built-in matrix element correction
is shown in magenta.
  In fact, the $y_3$ distribution is not very sensitive to
  the choices of parameters considered,
  particularly for smaller values of {\tt AFACT(1)},
  except near the matching point of $y_3=10^{-3}$.  Nonetheless, the
  choice of ${\frac{1}{2}k_T^2}$ as the evolution variable and
  ${\frac{1}{8}k_T^2}$ as the argument of $\alpha_S$ yields the
  best agreement of the choices shown.
  The agreement with \HW\ for the differential distribution
  $\frac1{\sigma}\frac{d\sigma}{d\log y_4}$ at parton level, Fig.\,\ref{fig:nllscalesB}, 
  depends much more on the choice of scales, with {\tt QFACT(1)=1/2} and
  {\tt AFACT(1)=1/8} giving the best results.
   These particular results are for
  a matching scale of $y=0.001$ which corresponds to a value of 
  ${\tt SCLCUT}=8.31\ \rm{GeV}^2$. This is a very low value for the
  matching scale and therefore the difference between the different choices
  is enhanced.   For either higher matching scales or 
  centre-of-mass energies the 
  differences are smaller.
  Using the
  \HW\ form factors (not shown here)
  gives worse agreement for these distributions.

  The previous plots used the scale choice {\tt ISCALE=1}.
  The dependence on the specific choice 
  is demonstrated in Fig.\,\ref{fig:nlliscale} for
  the  differential cross-section with respect to $y_3$.
  The definition in terms of the $k_T$-measure gives the
  best results, as we expected for the \HW\ (angular--ordered) algorithm. 
  All the remaining results use this choice.

  The effect of varying the minimum starting scale of the \HW\ parton shower,
  {\tt QFACT(3)}, 
  together with the variation of the scale of $\as$, on the parton-level
  differential cross-section with respect to $y_4$ is shown in Fig.\,\ref{fig:cut-off}.
  In $\ee$ collisions the main effect of this parameter is to 
  allow partons from the matrix element to produce more radiation, particularly
  those which are close to the cut-off in the matrix element.
  This tends to increase the smearing of the Durham jet 
  measure for these partons
  causing more events from the matrix element to
  migrate below the matching scale after the
  parton shower. Despite the cut-off on emission above the matching scale in the
  parton shower, some emissions occur above the matching scale in the parton
  shower, and this can help to ensure a smooth matching.
  The choice of {\tt QFACT(3)=4.0} gives the best results.

  So far, 
  we have considered the $y_3$ and $y_4$ distributions, which depend on the properties of the
  hardest one or two additional jets generated by either the matrix element or
  parton shower, and thus are not very dependent on the treatment
  of the highest multiplicity matrix element -- four additional jets in
  our numerical work.  Rather than study a higher order distribution, we
  focus still on the $y_3$ distribution but vary the matrix element
  multiplicity.  The rows of
  Fig.\,\ref{fig:eenjets} show the results of truncating the
  matrix element results $\epem\to q\bar q n$ for different $n=0,1,2,3$,
  while the columns show the dependence on the prefactor of the
  the scale in the argument for $\alpha_S$.  
In general the option {\tt IFINAL=3}
  gives the best results when only low multiplicity matrix elements are used.
  This corresponds to removing the Sudakov reweighting of the 
  external partons (between a cluster scale and $d_{\rm ini}$) and 
  performing the shower with a veto above the scale of the last emission in the
  matrix element.
  Also, both of the new prescriptions ({\tt IFINAL=2,3})
  perform better than the original \CK\ prescription if only low
  jet multiplicity matrix elements are used. As the number of jets in
  the highest multiplicity matrix element increases the differences
  between the prescriptions decreases, because the relative importance
  of this contribution is decreased.

  Up to this point,  we have discussed all the parameters relevant for the 
  simulation of
  $\ee\ra\rm{jets}$ in the \HW-\CK\ procedure 
  apart from the matching scale, $d_{\rm{ini}}$.
  In principle the results should be relatively insensitive to the choice of  
  this scale.  In practice, there is a dependence, because (1) we must
  truncate the matrix element calculation at some order, and (2) the
  parton shower may or may not give an adequate description of physics
  below the cut-off.  The effect of varying the matching scale on the
  differential cross-section with respect to $y_3$ is shown in 
  Fig.\,\ref{fig:eelepy3part} for $\sqrt{s}=M_Z$ and in
  Fig.\,\ref{fig:eenlcy3part} for $\sqrt{s}=500$~GeV at parton level.
  In general the agreement between \HW\ and the results of the \CK\ 
  algorithm is good, and improves as the matching scale is increased. 
  Similarly the agreement is better at $\sqrt{s}=500$~GeV than 
  $\sqrt{s}=M_Z$.

  Normally we would hope than the remaining differences at the matching scale would
  be smoothed out by the hadronization model. However as can be seen in 
  Fig.\,\ref{fig:eelepy3had} the \HW\ hadronization model distorts the 
  parton-level results and produces a double peaked structure for the
  differential cross section with respect to $y_3$ at hadron level.
  This problem is due to the treatment of events where there is no radiation
  in the parton shower. In these events the \HW\ hadronization model
  produces large mass clusters and their treatment is sensitive to the parameters
  of the hadronization model which control the splitting of these clusters. Hopefully
  retuning these parameters would improve the agreement for this distribution.
  The results at $\sqrt{s}=500$~GeV, shown in Fig.\ref{fig:eenlcy3had},
  where the fraction of events with these
  massive clusters is smaller, are much closer to the original \HW\ result.

  To summarize, the best results for the \HW-\CK\, matching in
  $\ee$ collisions are obtained using 
  NLL Sudakov form factors with a next-to-leading order 
  $\as$.\footnote{We have not discussed the choice of the order of $\as$ but 
                  NLO $\as$ is required as this is always used in \HW.}
  The best definition of the scale
  parameter is  {\tt ISCALE=1}, corresponding to $k_T^2$ as the
  evolution variable.   The effect of varying the prefactors of the
  scales is less dramatic and while 
  {\tt QFACT(1,2,3)=1/2} and {\tt AFACT(1,2)=1/4} are the best values
  these parameters can still varied in order to assess the effect of this
  variation on the results.
  The best choice for the treatment of the highest multiplicity matrix element is
  {\tt IFINAL=3} although provided sufficiently high multiplicity matrix elements
  are included the effect of this choice is small.
  Unfortunately, the hadron-level results are not as well-behaved as
the parton-level ones, but this may be resolved by retuning the
hadronization model.  Though it has not been thoroughly investigated,
this problem may be ameliorated when using \PY\ with this prescription.

\subsubsection{Pseudo-Shower Results}

  In this section, we show the results from the alternative scheme 
outlined in Sec.~\ref{sec:num} based on using the Sudakov form factors and
scale choices of the generators themselves.  
We show results in particular
for the \PY\ event generator, using the {\tt LUCLUS} measure for clustering
the partons to construct a parton-shower history.  
Clustering and matching is done on the {\tt LUCLUS} variable.
However, once a correction procedure has been applied, the
results can be used for any collider predictions -- one is not
limited to studying {\tt LUCLUS}-type variables for example.
To facilitate comparison, 
we show final-state results based on the $k_T$-clustering measure as in the
previous discussion.
The matrix element predictions are the same as those in the
previous analysis (which more strictly adheres to the \CK\ methodology),
but have been clustered using the {\tt LUCLUS} measure, so that they
constitute a subset of those events.

In the pseudo-shower approach, there are less free choices, so we
will limit ourselves to the final results.
Figure\,\ref{fig:ycut001}
shows the
differential $y_i$ distributions for cut-offs of $y_{\rm cut}=10^{-3}$
and  $10^{-2}$,
respectively, at the hadron level for $i=3,4$ and $5$.
Similar results (not shown) hold for $y_{\rm cut}=5\times 10^{-3}$.
The cut-off $d_{\rm ini}$ used for generating the matrix element
sample is shown as a vertical line in each of
the distributions; however, an additional 
offset $d_{0}$ was added to each cut-off to account for
any remaining mismatch between the parton-shower kinematics and
the final-state parton/hadron kinematics.  
Removing this offset induces a radiation dip at the cut-off scale.
A fixed value of $d_{0}=2$ GeV
was used for this study.  
The default \PY\ prediction including the
matrix element correction is shown as the dashed-line.  The
corrected distribution is the solid line, and is a sum of matrix element
predictions for $\epem\to$ 2 partons (red), 3 partons (green),
4 partons (blue), 5 partons (yellow) and 6 partons (magenta).
It is interesting to compare the relative contributions of a
given matrix element multiplicity for the different cut-offs.
Note that, unlike in the \CK-\HW\ procedure, there is a significant
overlap of the different contributions.  This is because the
different matrix element samples are clustered in a different
variable (the {\tt LUCLUS} measure) and then projected into
the $k_T$-measure.
The results generally agree with \PY\ where they should, and
constitute a more reliable prediction for the high-$y$ end
of the $y_i$ distributions.
For the lower cut-off, $10^{-3}$, the results are more sensitive to the
treatment of the highest multiplicity matrix element, as indicated
by the right tail of the matching predictions on the $y_4$ and $y_5$
distributions.  For the higher cut-off, $10^{-2}$, the actual contribution
of the 6 parton-matrix element is numerically insignificant, and there
is little improvement over the \PY\ result.
Results at the parton level (not shown) are similar.
Applying the same procedure to \HW\ 
(with the \HW\ choice of starting scales and 
argument for $\alpha_s$) yields similar results.

\subsection{Hadron-Hadron Collisions}

Up to this point, we have benchmarked two procedures for matching
matrix element calculations with parton showers -- the \HW-\CK\
procedure and the pseudo-shower procedure using \HW\ and \PY.  
 In this section,
we apply these methods to particle production at hadron colliders.
For the \HW-\CK\ procedure,
  many of the choices of parameters and 
  options were explored in the previous
  section.  Here we focus on the additional parameters which are relevant
  to hadron collisions, 
  although the effects of some of the parameters are different.
  In particular we have to make a choice of which variant
  of the $k_T$-algorithm to use.  For the pseudo-shower procedure,
  we apply the same method as for $\rm{e}^+\rm{e}^-$ 
  in a straight-forward fashion.

  In hadron collisions, since the center-of-mass energy of the
hard collision is not
transparent to an observer,  
  we study the differential distributions with
  respect to the square root of the $k_T^2$-measure defined in 
  Eqn.\,\ref{eqn:angle2}, 
  as this is related to the differential cross section
  with respect to the $p_T$ of the jet.

\subsubsection{\HW-\CK\ Results}

  In order to assess the effects of the varying of the scales for $\as$ and the
  minimum starting scale for the shower in hadron-hadron collisions we started
  by studying electroweak gauge boson
  production using the monotonic $p_T^2$-scheme and a
  matching scale $d_{\rm{ini}}=400~\rm{GeV}^2$
  at the Tevatron for a centre-of-mass energy of
  $1.96$~TeV. The $\rm{W}^++\rm{jets}$ events we are
  considering were generated including the leptonic decay of the W with
  no cuts on the decay leptons. The $\rm{Z}+\rm{jets}$ events were also 
  generated including the leptonic decay of the gauge boson and included
  the contribution of the photon exchange diagrams. To control the rise
  in the cross section at small invariant masses of the lepton pair, 
  $m_{\ell\ell}$,
  a cut was imposed requiring $m_{\ell\ell}\geq20$~GeV.

  The differential cross section with respect to $\sqrt{d_1}$ is shown in
  Fig.\,\ref{fig:Wd1a} for W~production and for Z~production
  in Fig.\,\ref{fig:Zd1a}. 
  The differential cross section with respect to $\sqrt{d_2}$  for 
  W~production is shown in Fig.\,\ref{fig:Wd2a}. 
  The results for $d_1$ for W~production show good agreement between \HW\
  and the \CK\ result for all the parameters, however the mismatch
  at the matching scale increases as {\tt QFACT(3,4)} increases. The is due 
  to the same effect we observed in $\ee$ collision, \ie more
  smearing of the one jet matrix element result causing more of these events
  to migrate below the matching scale.
  However the results for both Z~production and $d_2$ in W~production
  show a much larger discrepancy at the matching scale. In both
  cases there is a depletion of radiation below the matching scale with
  respect to the original \HW\ result. This is not seen in the
  $d_1$ distribution for W~production as here the initial-state
  parton shower always
  starts at the W~mass.\footnote{In practice this is smeared with the
  Breit-Wigner distribution due to the width of the W.}
  However the parton shower for Z~production often starts at the cut-off
  on the lepton pair mass and the parton shower of the $\rm{W}+1~\rm{jet}$
  matrix
  element often starts at a much lower scale. One possible 
  solution to this problem, at least for W~production, is to always 
  start the initial-state parton shower at the highest scale in the process.
  In Z~production however this cannot solve
  the problem of lack of radiation from events which have a low mass lepton pair.

  The best solution to this problem is to decouple the minimum starting
  scale for the initial- and final-state parton showers. In practice
  we want to use a low value for the final-state shower in order to 
  reduce the smearing of the higher multiplicity matrix elements and 
  a high value for the initial-state shower to avoid the problem of
  the radiation dip below the matching scale.

\subsubsection{Pseudo-Shower Results}

  In this section, we show the results of using the alternative scheme 
outlined in Sec.~\ref{sec:num}.
We show results 
for the \PY\ and \HW\ event generators applied to $W^+$ production
at the Tevatron, 
using the {\tt LUCLUS} measure for clustering 
the partons to construct a parton-shower history.  Again, since 
final results should not depend on the correction methodology, 
we show results based on the $k_T$-clustering measure as in the
previous discussion.  In all cases, we use a factorization scale
(which sets the upper scale for parton showers) equal to the
transverse mass of the $W^+$ boson: $Q_F=\sqrt{M_W^2+p_{TW}^2}=
\sqrt{E^2_W-p^2_{3W}}$.

Figure \,\ref{fig:W10sm} shows the
differential $k_T$-distributions for a cut-off of $10$ GeV,
using the pseudo-shower procedure with \PY\ to generate the parton shower.  
Similar results hold for cut-offs of 15 and 20 GeV used in this
study.
As for the $\epem$ case, these distributions are constructed from
fully hadronized events.  For the case of hadronic collisions, this requires
that partons from the underlying event and the hadronic remnants
of the beam particles are {\it not} included in
the correction.  In \PY\ and \HW, such partons can be identified
in a straight-forward manner.
The default \PY\ prediction including the
matrix element correction is shown as the dashed line.  The
corrected distribution is in black, and is a sum of matrix element
predictions for $p\bar p\to W+$ 0 partons (red), 1 parton (green),
2 partons (blue), 3 partons (yellow) and 4 partons (magenta).
The largest $k_T$-cluster value, which roughly corresponds to
the highest $p_T$ jet, agrees with the \PY\ result, but is 
about $33\%$ higher by $k_T=40$ GeV.  Such an increase is
reasonable, since matrix element corrections from two or more
hard partons is not included in \PY.  The deviations from
default \PY\ become greater when considering higher $k_T$-cluster
values: $k_T^3$ is roughly an order of magnitude larger in
the pseudo-shower method at $k_T=40$ GeV.  The transverse momentum
of the $W^+$ boson, however, is not significantly altered from
the \PY\ result, increasing by about $25\%$ at $k_T=40$ GeV.
Of course, larger deviations will be apparent at higher values
of transverse momentum.

The dependence on the cut-off is illustrated in
Figure \ref{fig:W345sm}, which shows the 
distributions for the 3rd, 4th, and 5th
largest value of $k_T$-cluster for the different choices of matching scale:
10 GeV (solid), 15 GeV (dashes), and 20 GeV (dots).
The default \PY\ prediction with the matrix element correction
(dot-dash) is shown for
comparison.  For $k_T^i>30$ GeV, the predictions are robust, though there is
a noticeable dependence with the matching scale for $k_T^i\le 30$ GeV.  The
$k_T^5$ distribution is generated purely from parton showering of the
other $W+n$ parton configurations.   
The largest variation among the predictions occurs around $k_T=20$ GeV,
corresponding to the largest cut-off, where the differential cross
section ranges over a factor of 3.  
Thus, for absolute predictions, the
choice of matching scale introduces a significant systematic bias.
On the other hand, most likely the data will be used to normalize distributions
with relatively loose cuts.  Figure \ref{fig:ratio} shows the ratio of the
distributions shown in Fig.~\ref{fig:W345sm} with respect to 
the distribution with a cut-off of 10 GeV, which exhibits far less variation
with the choice of matching scale.  
The ratio of the distribution for a cut-off of 15 GeV to 10 GeV is shown
(solid) and for 20 GeV to 10 GeV (dashed).
The most significant variation,
about 60\% for $k_T\sim 10-15$ GeV, occurs in
the ratio $k_T^3/k_T^5$, which is sensitive to the treatment of the highest
multiplicity matrix element ($W+4$ partons here), and would be expected to
show the greatest variation.  The variation is smaller for $k_T$ above 
the largest matching scale (20 GeV).

To test the whole pseudo-shower methodology, 
we now show results based on applying the
same algorithm with the \HW\ event generator.  A different starting scale
and argument for $\alpha_S$ is used, as noted previously.
The comparative $k_T$-distributions are shown in Figure~\ref{fig:W10sm_hw}.
Similar results hold for the cut-off values of 15 and 20 GeV used in
this study, and are not shown here for brevity.
The results are similar in nature to those from \PY, though the
spectra are are typically softer at the tail of the distributions.
Note that the pseudo-shower \HW\ results are compared to
\PY\ (not \HW) with the tuned underlying event model.  
Focusing attention on Figs.~\ref{fig:W345sm_hw}
and \ref{fig:ratio_hw}, which show the distributions $k_T^3, k_T^4$ and $k_T^5$
and their ratios,
we observe a smaller variation than for the \PY\ case.
The variation for $k_T^3/k_T^5$ in the range of 10 GeV is 
approximately 30\%.  
In general, the dependence on matching scale is smaller
than for the \PY\ results, both for the absolute shapes and ratios.

\section{Comparison to the MLM Approach}
\label{sec:mlm}
Recently, a less complicated method was suggested for adding parton
showers to $W+n$ parton matrix element calculations \cite{mlm}.
We denote this as the MLM method.  The resulting events samples were
meant for more limited applications, but it is worth commenting on
the overlap between the approaches.

The MLM method consists of several steps:
\begin{enumerate}
\item Generate $N_{\rm tot}$ events of uniform weight for $W+n$ partons at the 
tree level with cuts on $|\eta^i|<\eta^{\rm max}$, $E_T^i>E_T^{\rm min}$, 
and $\Delta R_{ij}>R^{\rm min}$,
where $i$ and $j$ denote partons.  The PDF's and $\alpha_s$ are
evaluated at the factorization scale $Q_F\sim M_W$ or $\sqrt{P_T^2+M_W^2}$.
The uniform weight of the events is the given by the total cross section
divided by the number of events: $\sigma/N_{\rm tot}$.
\item Apply a parton shower using \HW\ with a veto on $p_T>Q_F$, where
$p_T$ is the \HW\ approximation to the relative $p_T$ as described earlier.
By default, the starting scale for all parton showers is given by
$\sqrt{p_i\cdot p_j}$, where $i$ and $j$ are color--connected partons.
\item The showered partons are clustered into $N$ jets using a cone
algorithm with parameters $E_T^{\rm min}$ and $R^{\rm min}$.
If $N<n$, the event is reweighted by 0.  If $N\ge n$ (this
is the {\it inclusive} approach), the event is 
reweighted by 1 if each of the original $n$ partons is {\it uniquely}
contained within a reconstructed jet.  Otherwise, the event is reweighted
by 0.
\item At the end of the procedure, one is left with a sub-sample of the
original events with total cross section $\sigma N_{\rm acc}/N_{\rm tot}<\sigma$,
where $N_{\rm acc}$ are the number of events accepted (reweighted by 1).
\end{enumerate}

The method is well motivated.  It aims to prevent the parton shower from
generating a gluon emission that is harder than any emission already
contained in the ``hard'' matrix-element calculation.    The cuts on
$E_T$ and $\Delta R$ play the role of the clustering cuts on $k_T$
or $p_T$ (see Eq.~(\ref{eqn:angle2})).  Clearly, the same cuts applied
to the matrix element calculation are used to control the amount of
radiation from the parton shower.  However, a full clustering of the
event is not necessary, since no Sudakov form factors are applied
on internal lines, and \HW\ already has a default choice of starting
scales.  Based on the understanding of the numerical method for
applying the Sudakov reweighting, it is clear that the 
final step of rejecting emissions that are too hard
with respect to the matrix element calculation is the same as
applying a Sudakov form factor to the external lines only.
With respect to the internal lines, the reweighting coming
from an internal line in the parton-shower approach,
given by the product of the Sudakov form factor $\Delta(Q_{h},Q_{l})$
and the branching factor $\alpha_s(q_T)$, is to be compared to
the weight $\alpha_s(Q_F)$ in the MLM approach.  The size of
any numerical difference
between these factors is not obvious.  

To facilitate a direct comparison between the methods, we substitute the
cuts on $E_T$ and $\Delta R$ with a cut on the minimum $k_T$-cluster value
as in the original \CK\ proposal.  The jet-parton matching (step 3) is
further replaced by the requirement that the $(n+1)$st value of
$k_T$ after clustering the showered partons is less than the $n$th value
of $k_T$ from clustering the original partons, i.e $k_T^{n+1}<\tilde k_T^n$.
We will experiment with the choice of a veto on the parton shower, using
either $Q_F$ or $\tilde k_T^n$ for the internal \HW\ veto.

The resulting $k_T$-distributions are shown in Figure~\ref{fig:mlm_kt1}.
Similar results hold for the cut-off values of 15 and 20 GeV also used
in this study, but not shown here.
A comparison of the $k_T$-distributions
for different choices of matching scales is shown in Fig.~\ref{fig:mlm_kt345}.
While there are notable differences with the previous approaches, the
matching is nonetheless robust.  The $k_T^1$ distribution indicates some
enhancement-depletion of radiation above-below the matching scale, becoming more
noticeable as the matching scale is increased.  However, the $k_T^2$ distribution
does not suffer from this effect.  The variation with the matching scale 
is smaller than in the previous approaches, and is most noticeable for
$20<k_T<30$ GeV.

Finally, 
Fig.~\ref{fig:compare_ratio} shows a comparison of the $k_T$ ratios using
the pseudo-shower method with \HW\, the MLM method with \HW,
and the \HW-\CK\ method, all relative to the \PY\, pseudo-shower
result.
The cut-off used for this comparison is 15 GeV.  Below this scale, the
two \HW\ distributions are almost identical, and there is a significant
difference with \PY\ for distributions involving $k_T^5$ -- which is
generated by the parton shower.  Above the cut-off, the pseudo-shower
procedure and the MLM procedure are in close agreement, and are
on the order of 20\% higher than the \HW\ pseudo-shower results.  This,
then, is a good estimate in the range of uncertainty in these predictions.

\section{Discussion and Conclusions}
\label{sec:conclusions}

In this work, we report on our exploration of matching matrix element
predictions with parton showers using a methodology close to the
\CK\ algorithm suggested in   \cite{Catani:2001cc,Krauss:2002up}.
In sum, we have compared three different procedures: (1) 
a slightly expanded version of the \CK\ procedure using \HW\
as the parton-shower generator (but not limited in principal
to \HW) and exploiting the freedom to choose scales and cut-offs; 
(2) a version of the \CK\ procedure relying on pseudo-showers
and matched closely to the scales and cut-offs of \PY\ and \HW;
and (3) a much simpler procedure based on the approach of
M. Mangano.  All three of the procedures yield reasonable results.

The \HW-\CK\ procedure uses all of the elements of the original
\CK\ procedure, but expands upon them.  Several choices of
scale were investigated as starting points for the vetoed parton
shower, and a wide range of prefactors were explored as arguments
to the analytic NLL Sudakov form factor and $\alpha_S$.  The
variation of the results with these choices is shown in
the figures.  Optimized choices were settled upon based on
the smoothness of distributions, the agreement with \HW\ where
expected, and the apparent improvement over the default \HW\ 
predictions.   While this appears to be a tuning, the final
choices are easily justifiable.  Since \HW\ is an angular-ordered
shower, a variable such as $k_T$-cluster values is well suited as
a starting point for the \HW\ shower.  Because of the details 
of the \HW\ shower, a prefactor of $\frac{1}{2}$ for the scale
used in the Sudakov form factor is understandable, as well
as a prefactor of $\frac{1}{8}$ for the scale used in evaluating
$\alpha_S$.  The results presented are better at the parton level
than at the hadron level, which may require some tuning of the
\HW\ hadronization model.  These effects become less important
when considering scattering processes at higher energies or
when the cut-offs are larger.

The pseudo-shower procedure uses the Sudakov form factors of
\HW\ and \PY\ to numerically calculate the Sudakov suppression.
Since the Sudakov form factor is a probability distribution for
no parton emissions, the suppression factor can be determined
by starting showers from different stages of the parton-shower
history and discarding those events with emissions above 
a given cut-off.  Because of the nature of this approach, there
is less tuning of parameters.  To match the argument used in
$\alpha_S$ by default in \HW\ and \PY, a different clustering
scheme was used:  $p_T$ clustering or {\tt LUCLUS}-clustering.
Final results at the hadron level are shown in the figures.
In general, the hadron-level results are better than the
parton-level ones.  The use of {\tt LUCLUS} over {\tt KTCLUS}
was driven by the kinematics of the \PY\ shower.  We have
not checked whether {\tt KTCLUS} works as well or better for
the \HW\ results, and we leave this for future investigation.
We should also investigate the advantages of using the
{\it exact} clustering scheme of the individual generators:
invariant mass and angular ordering for \PY\ or just
angular ordering for \HW.  Also, since this work began, a
new model of final-state showering was developed for \PY\ which
is exactly of the {\tt LUCLUS} type.  This should also be
tested, and ideally the $p_T$-ordered shower could be expanded
to include initial-state radiation.  This is beyond the stated
aims of this work, which was to investigate the use of
\HW\ and \PY\ with minor modifications.

The MLM procedure is a logical extension of the procedure develop
by M. Mangano for adding partons showers to $W$+multijet events.
It entails $k_T$-clustering the parton-level events, adding
a parton shower (with \HW\ in practice, but not limited to it), and rejecting
those events where the parton shower generates a harder emission
(in the $k_T$-measure) than the original events.  This
approach yields a matching which is almost as good as the more complicated
procedures based on the \CK\ procedures explored
in this work.  The reason is not a pure
numerical accident.  The MLM procedure rejects events (equivalently,
reweights them to zero weight) when the parton shower generates
an emission harder than the lowest $k_T$ value of the given
kinematic configuration.  This is equivalent to the first
step of the pseudo-shower procedure
in the calculation of the Sudakov suppression when applied to the highest
multiplicity matrix element.  The remaining difference
is in the treatment of the internal Sudakov form factors 
and the argument of $\alpha_S$.  The agreement between the
pseudo-shower and MLM procedures implies that the product of
Sudakov form factors on internal lines with the factors of
$\alpha_S$ evaluated at the clustering scale is numerically
equivalent to the product of $\alpha_S$ factors evaluated 
at the hard scale.  It is worth noting that, for
the process at hand, $q\bar q'\to W+X$, only two of the
cluster values can be very close to the cut-off, and thus
only two of the $\alpha_S(k_T)$ values can be very large.
Also, at the matching scales considered in this study,
$10-20$ GeV, with a factorization scale on the order 
of $M_W$, $Q_F=\sqrt{M_W^2+P_{TW}^2}$,
a fixed order expansion is of similar numerical
reliability as the ``all-orders'' expansion of a resummation
calculation.  In fact, the resummation (parton shower) 
expansion is ideally suited for $Q\ll M_W$, whereas the
fixed order expansion is best applied for $Q\sim M_W$.
The matching scales used in this study
straddle these extremes.

Based on the study of these three procedures, we can make
several statements on the reliability of predicting
the shapes and rates of multijet processes at collider energies.
\begin{enumerate}
\item The three matching procedures studied here can be recommended.
They are robust to variation of the cut-off scale.
\item The relative distributions in $k_T$, for example, are
reliably predicted.  
\item The variation in the relative distributions from the three
procedures depends on the variable.  For variables within the
range of the matrix elements calculated, the variation is 20\%.
For variables outside this range, which depend on the
truncation of the matrix element calculation, the variation is
larger 50\%.  Of course, it is important to study the experimental
observables to correctly judge the senstivity to the cut-off and
methodology of matching.

\item More study is needed to determine the best method for
 treating the highest multiplicity matrix-element contributions.

\item The subject of matching is far from exhausted.  The procedures
presented here yield an improvement over previous matching prescriptions.
However, these methodologies are an {\it interpolation} procedure.
\end{enumerate}

\section*{Acknowledgments}

  PR thanks S. Gieseke, F. Krauss, M. Mangano, M. Seymour, B. Webber
and members of the Cambridge SUSY working group
for many useful discussions about the \CK\ algorithm.
SM thanks T. Sj\"{o}strand for similar discussions.
PR and SM both thank J. Huston for discussion, encouragement
and for asking the difficult questions.
This work was started at the Durham Monte Carlo workshop and presented at
the Durham 'Exotics at Hadron Colliders', Fermilab Monte Carlo,
Les Houches and CERN MC4LHC workshops. 
PR would like to thank the organisers of these workshops and 
the Fermilab
theory and computing divisions for their hospitality during this work.  
SM thanks the Stephen Wolbers and the Fermilab Computing Division
for access to the Fixed Target Computing Farms.
PR gives similar thanks to the support term at the Rutherford lab CSF facility.

\appendix

\bibliography{meps}

\newpage

\FIGURE[!ht]{
{
\begin{picture}(300,200)
\SetOffset(0,-50)
\ArrowLine(0,200)(50,150)\ArrowLine(50,150)(0,100)
\Text(0,200)[rc]{$\rm{e}^-$}
\Text(0,100)[rc]{$\rm{e}^+$}
\Photon(50,150)(100,150){5}{5}
\Text(75,145)[ct]{$\gamma/\rm{Z}$}
\Text(100,145)[ct]{\Red{$d_1$}}
\ArrowLine(100,150)(150,200)\ArrowLine(150,200)(200,250)
\Text(150,200)[br]{\Red{$d_3$}}
\Text(200,250)[cl]{\Red{$d_{\rm{ini}}$}, $q_1$}
\ArrowLine(150,100)(100,150)\ArrowLine(200,50)(150,100)
\Text(200,50)[cl]{\Red{$d_{\rm{ini}}$}, $\bar{q}_1$}
\Text(150,100)[tr]{\Red{$d_2$}}
\Gluon(150,200)(200,200){-5}{4}
\Gluon(200,200)(250,250){5}{5}
\Gluon(200,200)(250,150){5}{5}
\Text(250,250)[cl]{\Red{$d_{\rm{ini}}$}, $g_1$}
\Text(200,193)[tr]{\Red{$d_4$}}
\ArrowLine(250,150)(300,200)\ArrowLine(300,100)(250,150)
\Text(300,100)[cl]{\Red{$d_{\rm{ini}}$}, $\bar{q}_2$}
\Text(300,200)[cl]{\Red{$d_{\rm{ini}}$}, $q_2$}
\Text(250,150)[tr]{\Red{$d_5$}}
\Gluon(150,100)(250,100){5}{8}
\Text(250,100)[cl]{\Red{$d_{\rm{ini}}$}, $g_2$}
\end{picture}}
\caption{Example of the clustering of an $\rm{e}^+\rm{e}^-$ event. The values of
         the $k_T$-parameter at the nodes are such that 
         $d_{\rm{ini}}<d_5<d_4<d_3<d_2<d_1$. The parton shower of the quark
         $q_1$ starts at the scale $d_1$, as does that of the antiquark $\bar{q}_1$.
         The parton shower of the gluon $g_2$ starts at the scale $d_2$. The situation
         is more complex for the remaining gluon and quark-antiquark pair. The
         shower of the $q_2$ and $\bar{q}_2$ should start at scale the virtual gluon which 
         branched to produce them was produced. If the virtual gluon is harder than the
         gluon $g_1$, 
         this means that parton shower of $q_2$ and $\bar{q}_2$ starts at the
         scale $d_3$ while that of $g_1$ starts at $d_4$. However if the
         virtual gluon is softer than $g_1$ the parton shower of the  $q_2$ and
         $\bar{q}_2$ start at the scale $d_4$ while the parton shower of $g_1$
         starts at the scale $d_3$.}
\label{fig:e+e-eg}}

\FIGURE[!ht]{ 
\begin{picture}(300,200)
\SetOffset(0,-50)
\Photon(150,150)(200,150){5}{5}
\Text(175,145)[ct]{$W^-$}
\Text(150,145)[ct]{\Red{$d_1$}}
\ArrowLine(150,150)(50,250)\ArrowLine(50,50)(150,150)
\Text(90,200)[br]{\Red{$d_3$}}
\Text(90,100)[br]{\Red{$d_2$}}
\Text(10,250)[cl]{\Red{$d_{\rm{ini}}$}, $q_1$}
\Text(10,50)[cl]{\Red{$d_{\rm{ini}}$}, $\bar{q}_1$}
\Gluon(100,200)(200,200){-5}{8}
\Gluon(200,200)(250,250){5}{5}
\Gluon(200,200)(250,150){5}{5}
\Text(250,250)[cl]{\Red{$d_{\rm{ini}}$}, $g_1$}
\Text(200,193)[tr]{\Red{$d_4$}}
\ArrowLine(250,150)(300,200)\ArrowLine(300,100)(250,150)
\Text(300,100)[cl]{\Red{$d_{\rm{ini}}$}, $\bar{q}_2$}
\Text(300,200)[cl]{\Red{$d_{\rm{ini}}$}, $q_2$}
\Text(250,150)[tr]{\Red{$d_5$}}
\Gluon(100,100)(200,100){5}{8}
\Text(200,100)[cl]{\Red{$d_{\rm{ini}}$}, $g_2$}
\end{picture}
\caption{Example of the clustering of a $\rm{W}+\rm{jets}$ event. The values of
         the $k_T$-parameter at the nodes are such that 
         $d_{\rm{ini}}<d_5<d_4<d_3<d_2<d_1$. The parton showers of the 
         incoming quark $q_1$ and
          antiquark $\bar q_1$ start at the scale $d_1$ at 
         which they annihilated. The
         parton showers of the quarks ${q}_2$ and $\bar{q}_2$ start at the
         scale $d_4$ at which the virtual gluon they came from was produced,
	 assuming that this gluon is softer than the gluon, $g_1$.
         The parton shower of the gluon $g_2$ starts at the scale $d_2$, 
         and the shower of gluon $g_1$ starts at the scale $d_3$
         at which the gluon which branched to produce it
         was produced. }
\label{fig:W+jetseg}}

\FIGURE[!ht]{
\includegraphics[width=0.55\textwidth,angle=90]{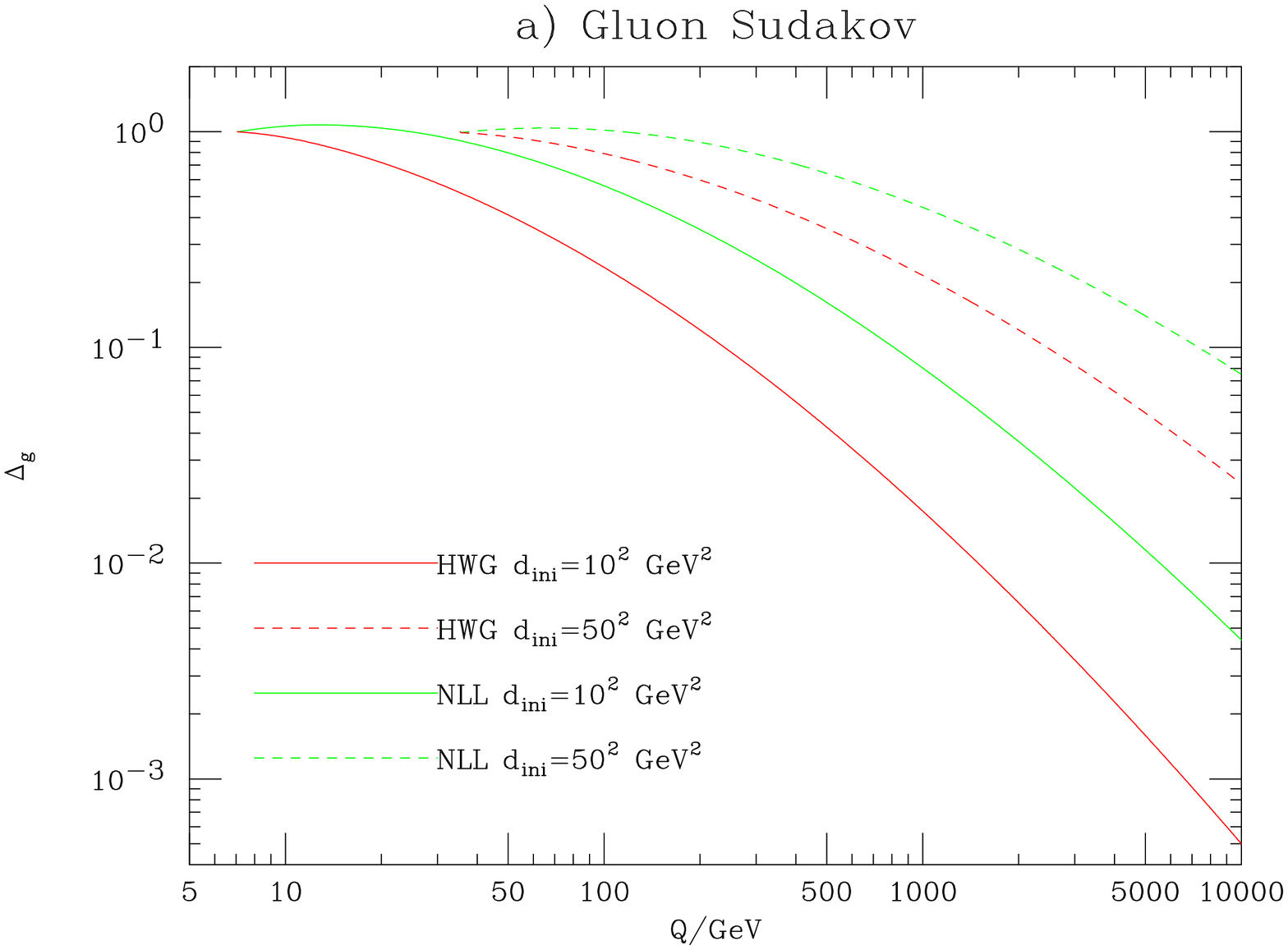}\hfill
\includegraphics[width=0.55\textwidth,angle=90]{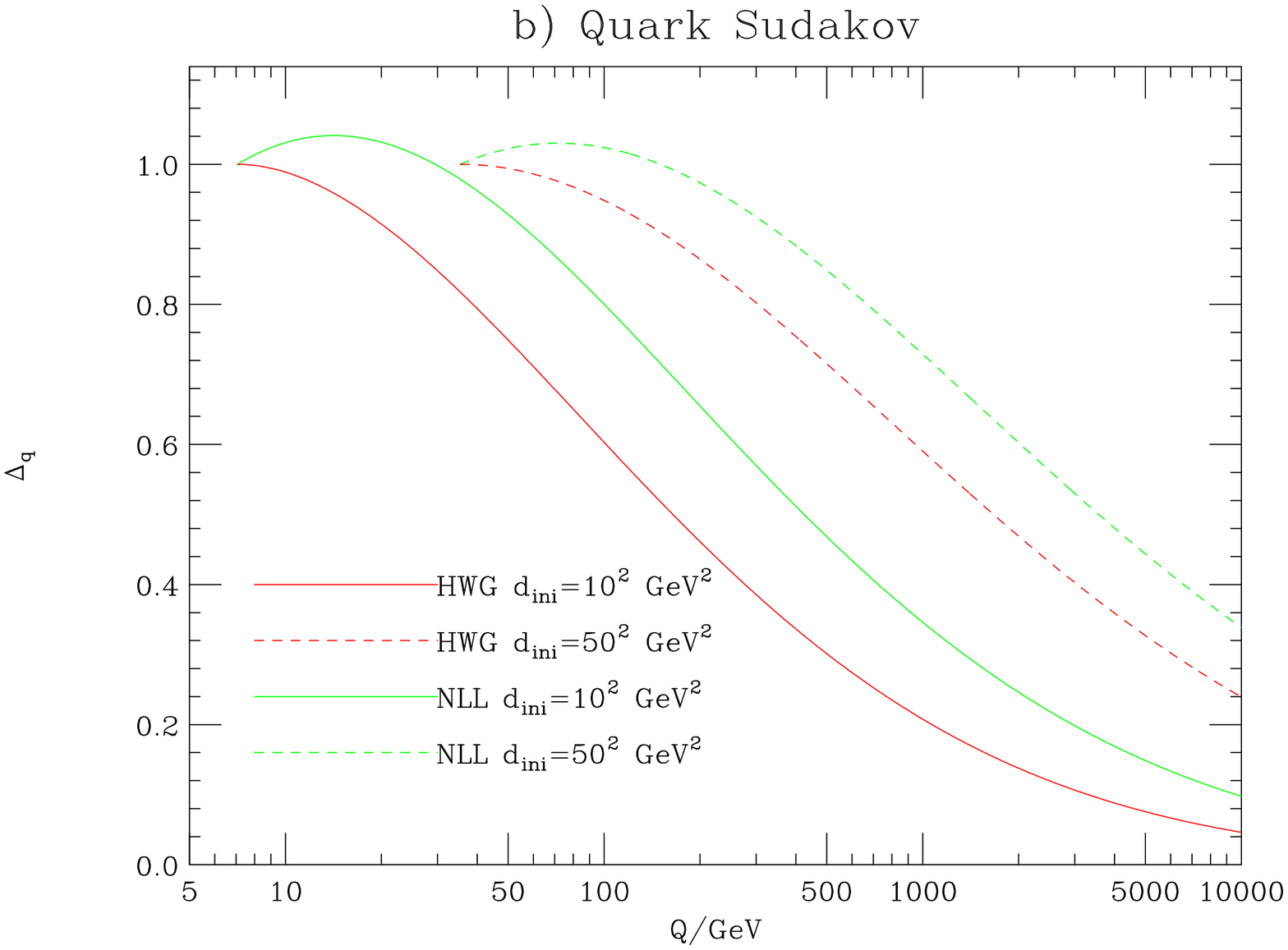}
\caption{\HW\ and NLL gluon $\Delta_g$ and quark $\Delta_q$
         Sudakov form factors for 
        \mbox{$d_{\rm{ini}}=10^2,\ 50^2\ \rm{GeV}^2$}. In both cases 
        the \HW\ NLO $\alpha_S$ has been used.  The form factor 
        represents the probability that a given species of parton
        will evolve from the high-scale (x-axis) to the cut-off 
        scale with no radiation.}
\label{fig:sudakov}}

\FIGURE[!ht]{
\begin{picture}(400,400)
\put(50,0){\includegraphics[bbllx=100,bblly=0,bburx=550,bbury=550,angle=90]{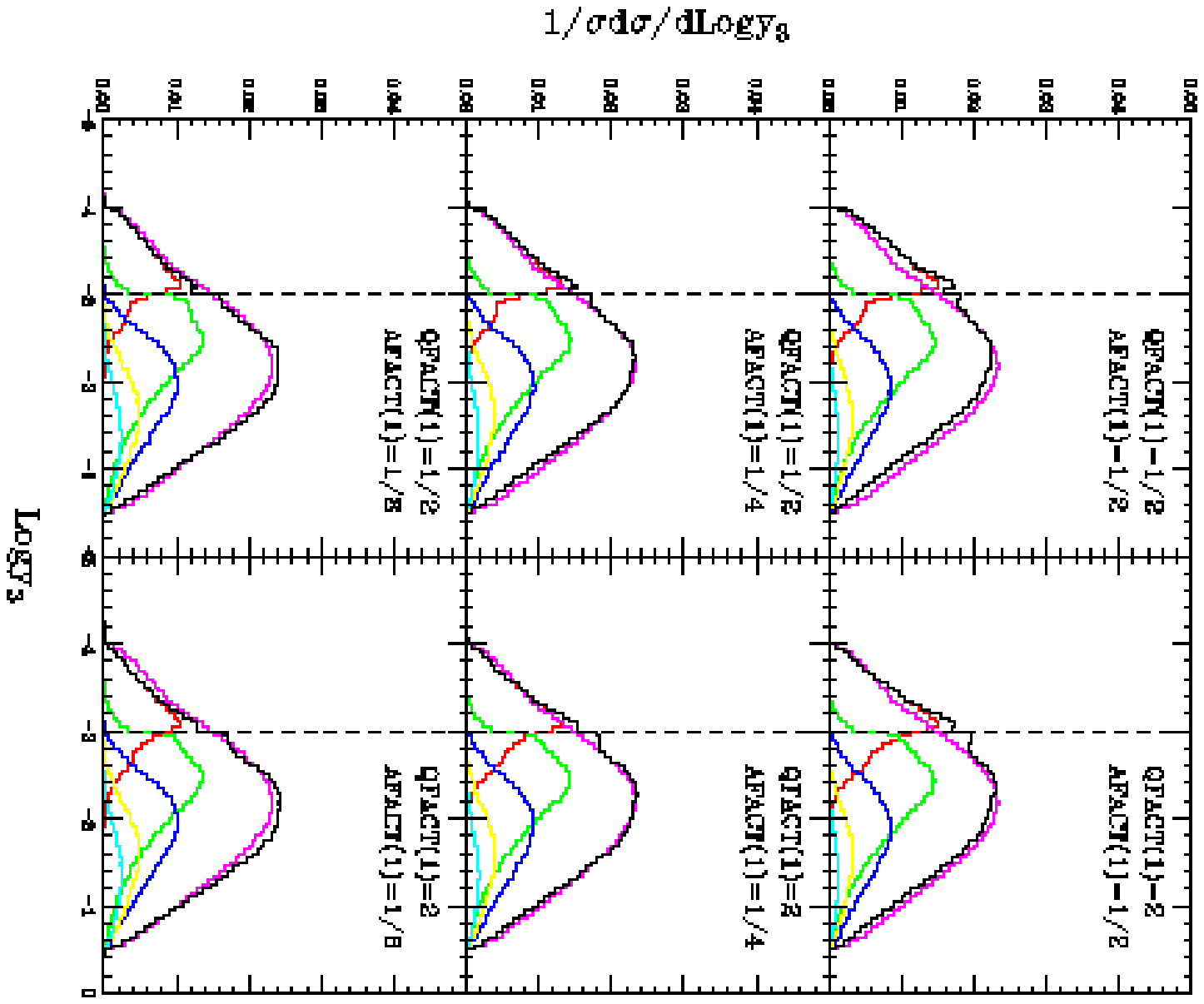}}
\Text(0,375)[rc]{\black ME}
\Text(0,350)[rc]{\magenta HW}
\Text(0,325)[rc]{\red 2 jets}
\Text(0,300)[rc]{\blue 3 jets}
\Text(0,275)[rc]{\green 4 jets}
\Text(0,250)[rc]{\yellow 5 jets}
\Text(0,225)[rc]{\cyan 6 jets}
\put(125,100){\Large \red $\star$}
\end{picture}
\caption{Effect of varying the prefactors for the scale in the Sudakov form factors
         and $\alpha_S$ using NLL Sudakov form factors
         on the parton-level 
         differential cross section $\frac{1}{\sigma}\frac{d\sigma}{d\log y_3}$ in 
         $\ee$ collisions at $\sqrt{s}=M_Z$. The parameters
         were set so that {\tt QFACT(1)=QFACT(2)} and {\tt AFACT(1)=AFACT(2)}.
         The default result of \HW\ is shown as a magenta line, 
         the result of the \CK\ algorithm is shown as a black line.
         The contribution to the \CK\ result of the different jet multiplicities
         are also shown, red is the 2 jet component, green is the 3 jet component,
         blue is the 4 jet component, yellow is the 5 jet component and cyan is the
         6 jet component. These results are for
         a matching scale of $y=0.001$, shown as a vertical dashed line,
          which corresponds to a value of 
         ${\tt SCLCUT}=8.31\ \rm{GeV}^2$ and use the original \CK\
         treatment of the highest multiplicity matrix element.}
\label{fig:nllscales}}

\FIGURE[!ht]{
\begin{picture}(400,400)
\put(50,0){\includegraphics[bbllx=100,bblly=0,bburx=550,bbury=550,angle=90]{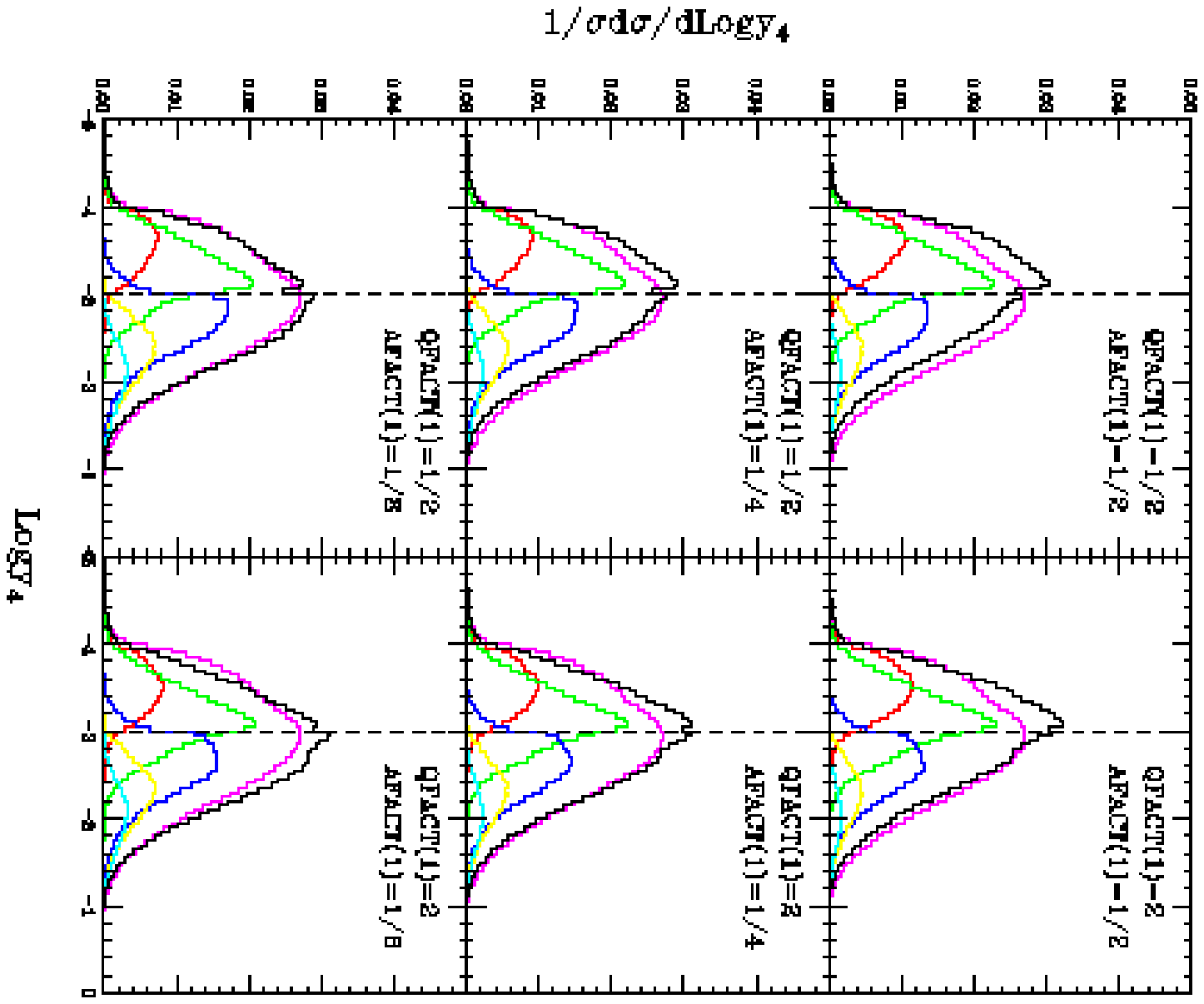}}
\Text(0,375)[rc]{\black ME}
\Text(0,350)[rc]{\magenta HW}
\Text(0,325)[rc]{\red 2 jets}
\Text(0,300)[rc]{\blue 3 jets}
\Text(0,275)[rc]{\green 4 jets}
\Text(0,250)[rc]{\yellow 5 jets}
\Text(0,225)[rc]{\cyan 6 jets}
\put(125,100){\Large \red $\star$}
\end{picture}
\caption{Effect of varying the prefactors for the scale in the Sudakov form factors
         and $\alpha_S$ using NLL Sudakov form factors
         on the  parton-level 
         differential cross section $\frac{1}{\sigma}\frac{d\sigma}{d\log y_4}$ in 
         $\ee$ collisions at $\sqrt{s}=M_Z$. The parameters
         were set so that {\tt QFACT(1)=QFACT(2)} and {\tt AFACT(1)=AFACT(2)}.
         The default result of \HW\ is shown as a magenta line, 
         the result of the \CK\ algorithm is shown as a black line.
         The contribution to the \CK\ result of the different jet multiplicities
         are also shown, red is the 2 jet component, green is the 3 jet component,
         blue is the 4 jet component, yellow is the 5 jet component and cyan is the
         6 jet component. These results are for
         a matching scale of $y=0.001$, shown as a vertical dashed line,
          which corresponds to a value of 
         ${\tt SCLCUT}=8.31\ \rm{GeV}^2$ and use the original \CK\
         treatment of the highest multiplicity matrix element.}
\label{fig:nllscalesB}}

\FIGURE[!ht]{
\begin{picture}(400,400)
\put(50,0){\includegraphics[bbllx=100,bblly=0,bburx=550,bbury=550,angle=90]{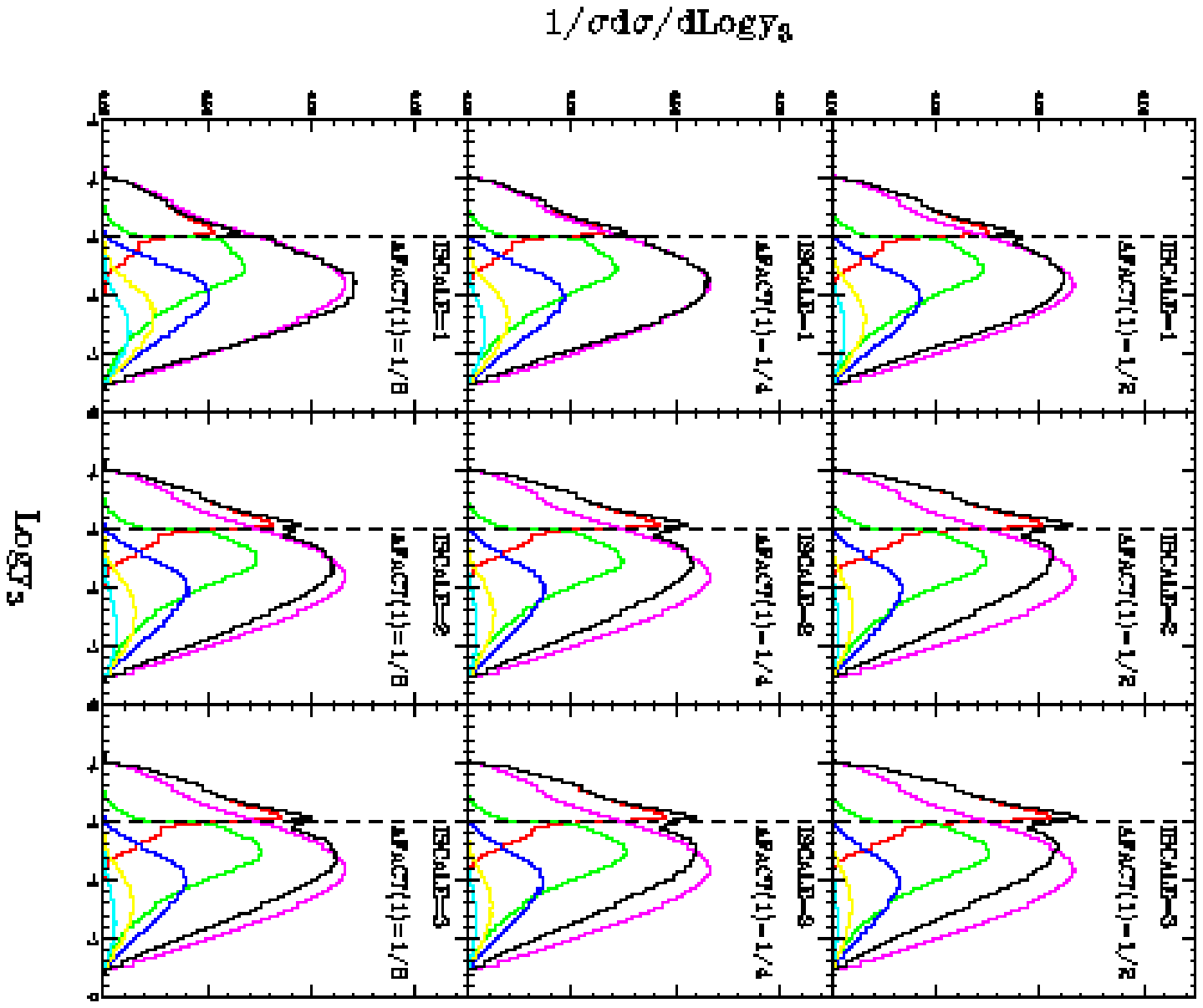}}
\Text(0,375)[rc]{\black ME}
\Text(0,350)[rc]{\magenta HW}
\Text(0,325)[rc]{\red 2 jets}
\Text(0,300)[rc]{\blue 3 jets}
\Text(0,275)[rc]{\green 4 jets}
\Text(0,250)[rc]{\yellow 5 jets}
\Text(0,225)[rc]{\cyan 6 jets}
\put(125,100){\Large \red $\star$}
\end{picture}
\caption{Effect of varying the choice of scale in the Sudakov form factors
         and $\alpha_S$ using NLL Sudakov form factors
         on the  parton-level 
         differential cross section $\frac{1}{\sigma}\frac{d\sigma}{d\log y_3}$
         in 
         $\ee$ collisions at $\sqrt{s}=M_Z$. The parameters
         were set so that {\tt QFACT(1)=QFACT(2)=1/2} and {\tt AFACT(1)=AFACT(2)}.
         The default result of \HW\ is shown as a magenta line, 
         the result of the \CK\ algorithm is shown as a black line.
         The contribution to the \CK\ result of the different jet multiplicities
         are also shown, red is the 2 jet component, green is the 3 jet component,
         blue is the 4 jet component, yellow is the 5 jet component and cyan is the
         6 jet component. These results are for
         a matching scale of $y=0.001$, shown as a vertical dashed line,
          which corresponds to a value of 
         ${\tt SCLCUT}=8.31\ \rm{GeV}^2$ and use the original \CK\
         treatment of the highest multiplicity matrix element.}
\label{fig:nlliscale}}

\FIGURE[!ht]{
\begin{picture}(400,400)
\put(50,0){\includegraphics[bbllx=100,bblly=0,bburx=550,bbury=550,angle=90]{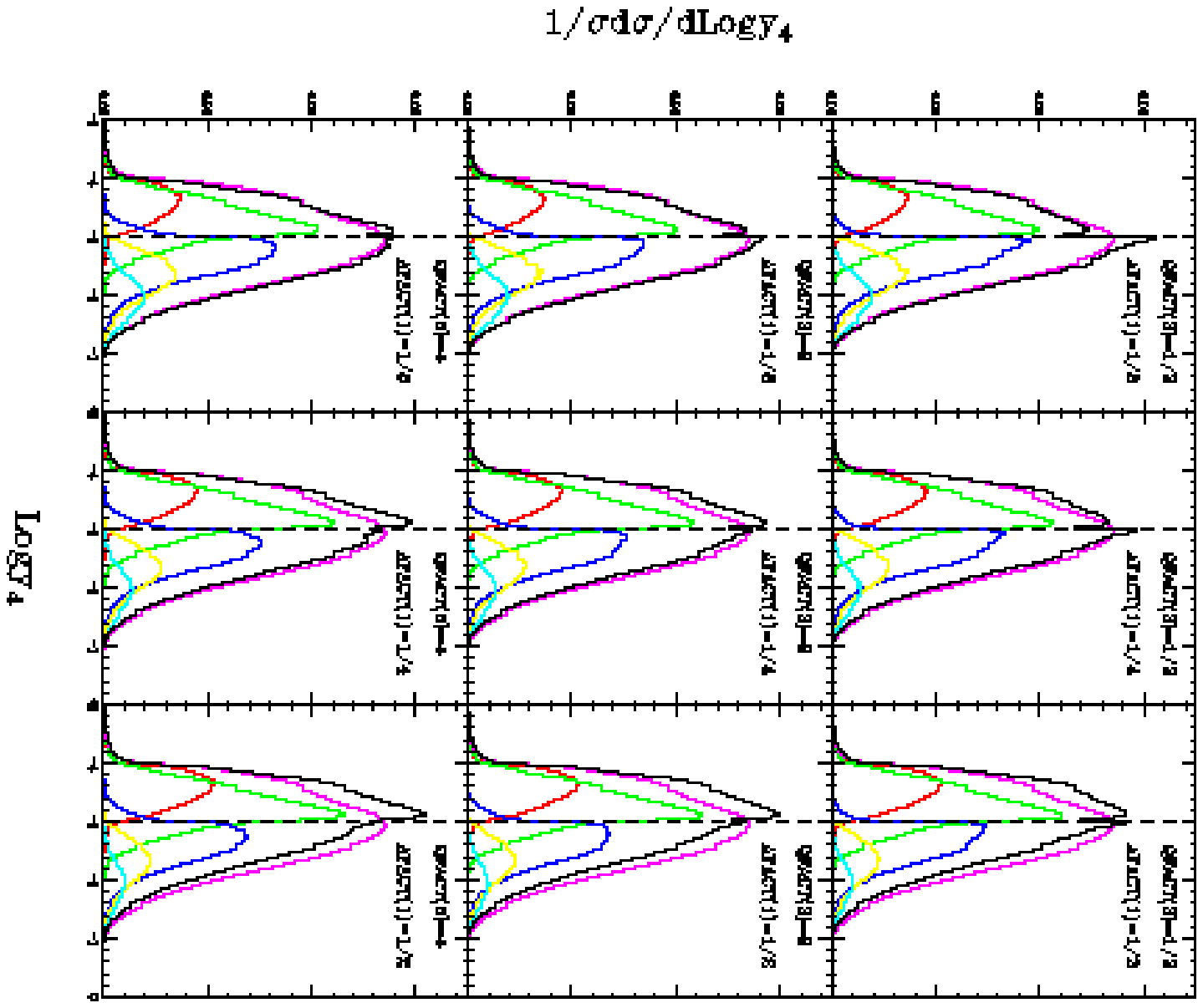}}
\Text(0,375)[rc]{\black ME}
\Text(0,350)[rc]{\magenta HW}
\Text(0,325)[rc]{\red 2 jets}
\Text(0,300)[rc]{\blue 3 jets}
\Text(0,275)[rc]{\green 4 jets}
\Text(0,250)[rc]{\yellow 5 jets}
\Text(0,225)[rc]{\cyan 6 jets}
\put(100,100){\Large \red $\star$}
\end{picture}
\caption{Effect of varying the minimum starting scale of the parton shower and the scale
         of $\as$ using NLL Sudakov form factors
         on the  parton-level 
         differential cross section $\frac{1}{\sigma}\frac{d\sigma}{d\log y_4}$ in 
         $\ee$ collisions at $\sqrt{s}=M_Z$. The parameters
         were set so that {\tt QFACT(1)=QFACT(2)=1/2} and {\tt AFACT(1)=AFACT(2)}.
         The default result of \HW\ is shown as a magenta line, 
         the result of the \CK\ algorithm is shown as a black line.
         The contribution to the \CK\ result of the different jet multiplicities
         are also shown, red is the 2 jet component, green is the 3 jet component,
         blue is the 4 jet component, yellow is the 5 jet component and cyan is the
         6 jet component. These results are for
         a matching scale of $y=0.001$, shown as a vertical dashed line,
          which corresponds to a value of 
         ${\tt SCLCUT}=8.31\ \rm{GeV}^2$ and use the original \CK\
         treatment of the highest multiplicity matrix element.}
\label{fig:cut-off}}

\FIGURE[!ht]{
\begin{picture}(400,400)
\put(50,0){\includegraphics[bbllx=100,bblly=0,bburx=550,bbury=550,angle=90]{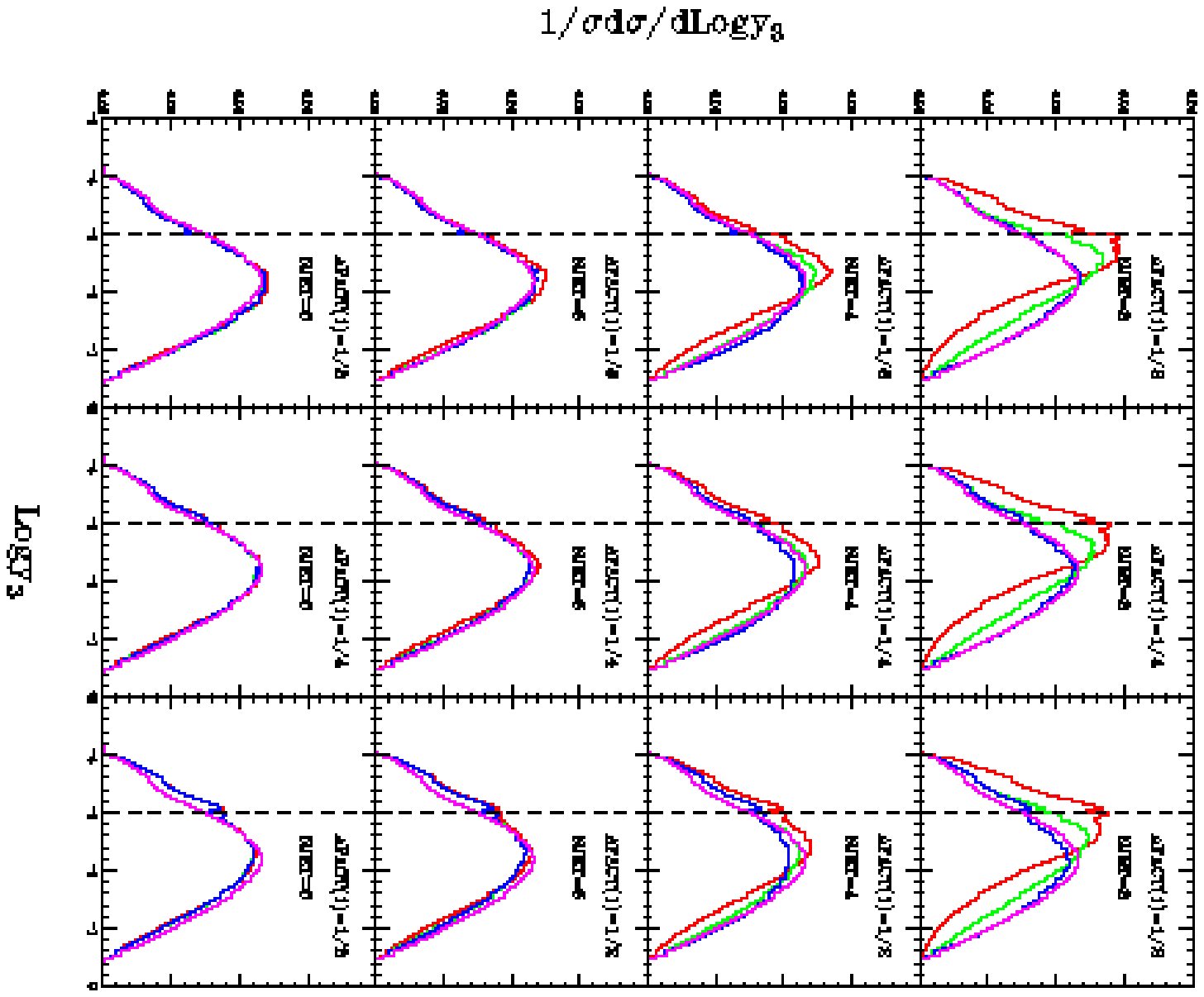}}
\Text(0,375)[lc]{\magenta HW}
\Text(0,350)[lc]{\red {\tt IFINAL=1}}
\Text(0,325)[lc]{\green {\tt IFINAL=2}}
\Text(0,300)[lc]{\blue {\tt IFINAL=3}}
\end{picture}

\caption{Effect of varying the treatment of the highest multiplicity matrix
         element using NLL Sudakov form factors
         on the  parton-level 
         differential
         cross section $\frac{1}{\sigma}\frac{d\sigma}{d\log y_3}$ in 
         $\ee$ collisions at $\sqrt{s}=M_Z$. The parameters
         were set so that {\tt QFACT(1)=QFACT(2)=1/2} and {\tt AFACT(1)=AFACT(2)}.
         The default result of \HW\ is shown as a magenta line, 
         the result of the \CK\ algorithm is shown as a red line for {\tt IFINAL=1},
         as a green line for {\tt IFINAL=2} and as a blue line for
         {\tt IFINAL=3}. These results are for
         a matching scale of $y=0.001$, shown as a vertical dashed line,
          which corresponds to a value of 
         ${\tt SCLCUT}=8.31\ \rm{GeV}^2$.}
\label{fig:eenjets}}

\FIGURE[!ht]{
\begin{picture}(400,400)
\put(50,0){\includegraphics[bbllx=100,bblly=0,bburx=550,bbury=550,angle=90]{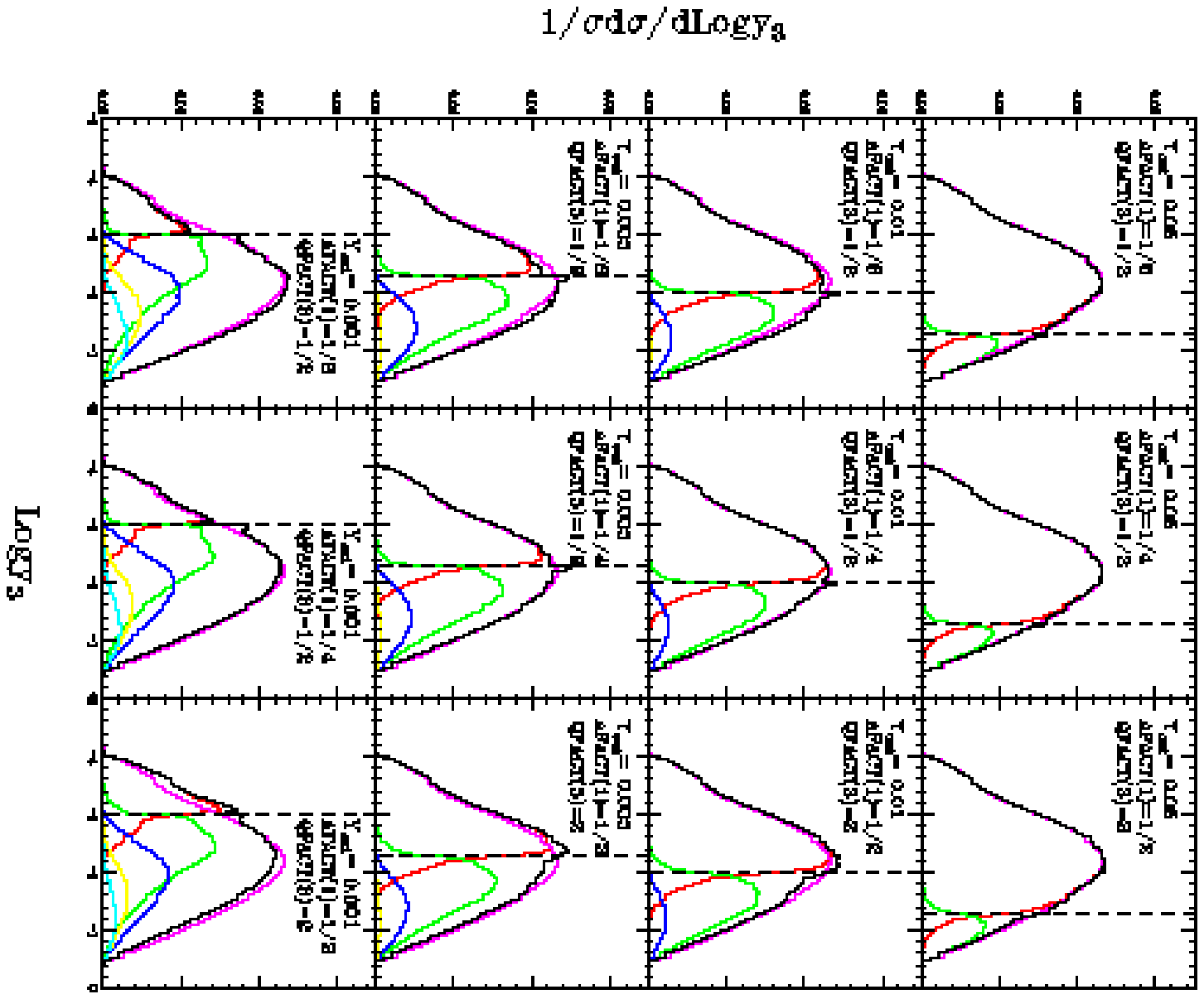}}
\Text(0,375)[rc]{\black ME}
\Text(0,350)[rc]{\magenta HW}
\Text(0,325)[rc]{\red 2 jets}
\Text(0,300)[rc]{\blue 3 jets}
\Text(0,275)[rc]{\green 4 jets}
\Text(0,250)[rc]{\yellow 5 jets}
\Text(0,225)[rc]{\cyan 6 jets}
\end{picture}

\caption{Effect of varying the matching scale
         using NLL Sudakov form factors
         on the  parton-level 
         differential
         cross section $\frac{1}{\sigma}\frac{d\sigma}{d\log y_3}$ in 
         $\ee$ collisions at $\sqrt{s}=M_Z$. The parameters
         were set so that {\tt QFACT(1)=QFACT(2)=1/2} and {\tt AFACT(1)=AFACT(2)}.
         The default result of \HW\ is shown as a magenta line, 
         the result of the \CK\ algorithm is shown as a black line.
         The contribution to the \CK\ result of the different jet multiplicities
         are also shown, red is the 2 jet component, green is the 3 jet component,
         blue is the 4 jet component, yellow is the 5 jet component and cyan is the
         6 jet component. The matching scale is shown as a vertical dashed line.
         The {\tt IFINAL=3} option was used for the highest multiplicity matrix
         element.}
\label{fig:eelepy3part}}
 
\FIGURE[!ht]{
\begin{picture}(400,400)
\put(50,0){\includegraphics[bbllx=100,bblly=0,bburx=550,bbury=550,angle=90]{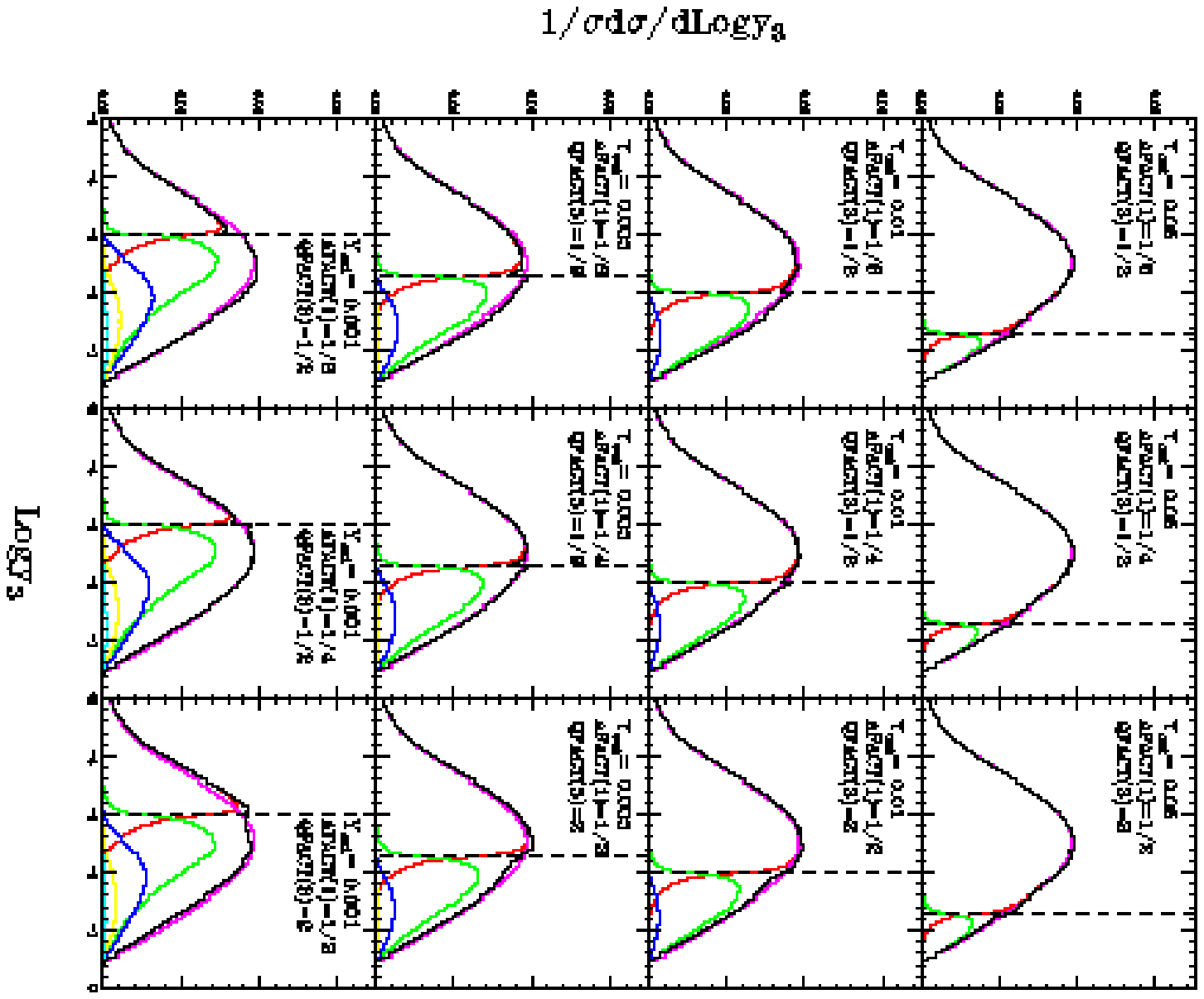}}
\Text(0,375)[rc]{\black ME}
\Text(0,350)[rc]{\magenta HW}
\Text(0,325)[rc]{\red 2 jets}
\Text(0,300)[rc]{\blue 3 jets}
\Text(0,275)[rc]{\green 4 jets}
\Text(0,250)[rc]{\yellow 5 jets}
\Text(0,225)[rc]{\cyan 6 jets}
\end{picture}

\caption{Effect of varying the matching scale
         using NLL Sudakov form factors
         on the  parton-level 
         differential
         cross section $\frac{1}{\sigma}\frac{d\sigma}{d\log y_3}$ in 
         $\ee$ collisions at $\sqrt{s}=500$~GeV. The parameters
         were set so that {\tt QFACT(1)=QFACT(2)=1/2} and {\tt AFACT(1)=AFACT(2)}.
         The default result of \HW\ is shown as a magenta line, 
         the result of the \CK\ algorithm is shown as a black line.
         The contribution to the \CK\ result of the different jet multiplicities
         are also shown, red is the 2 jet component, green is the 3 jet component,
         blue is the 4 jet component, yellow is the 5 jet component and cyan is the
         6 jet component. The matching scale is shown as a vertical dashed line.
         The {\tt IFINAL=3} option was used for the highest multiplicity matrix
         element.}
\label{fig:eenlcy3part}}

\FIGURE[!ht]{
\begin{picture}(400,400)
\put(50,0){\includegraphics[bbllx=100,bblly=0,bburx=550,bbury=550,angle=90]{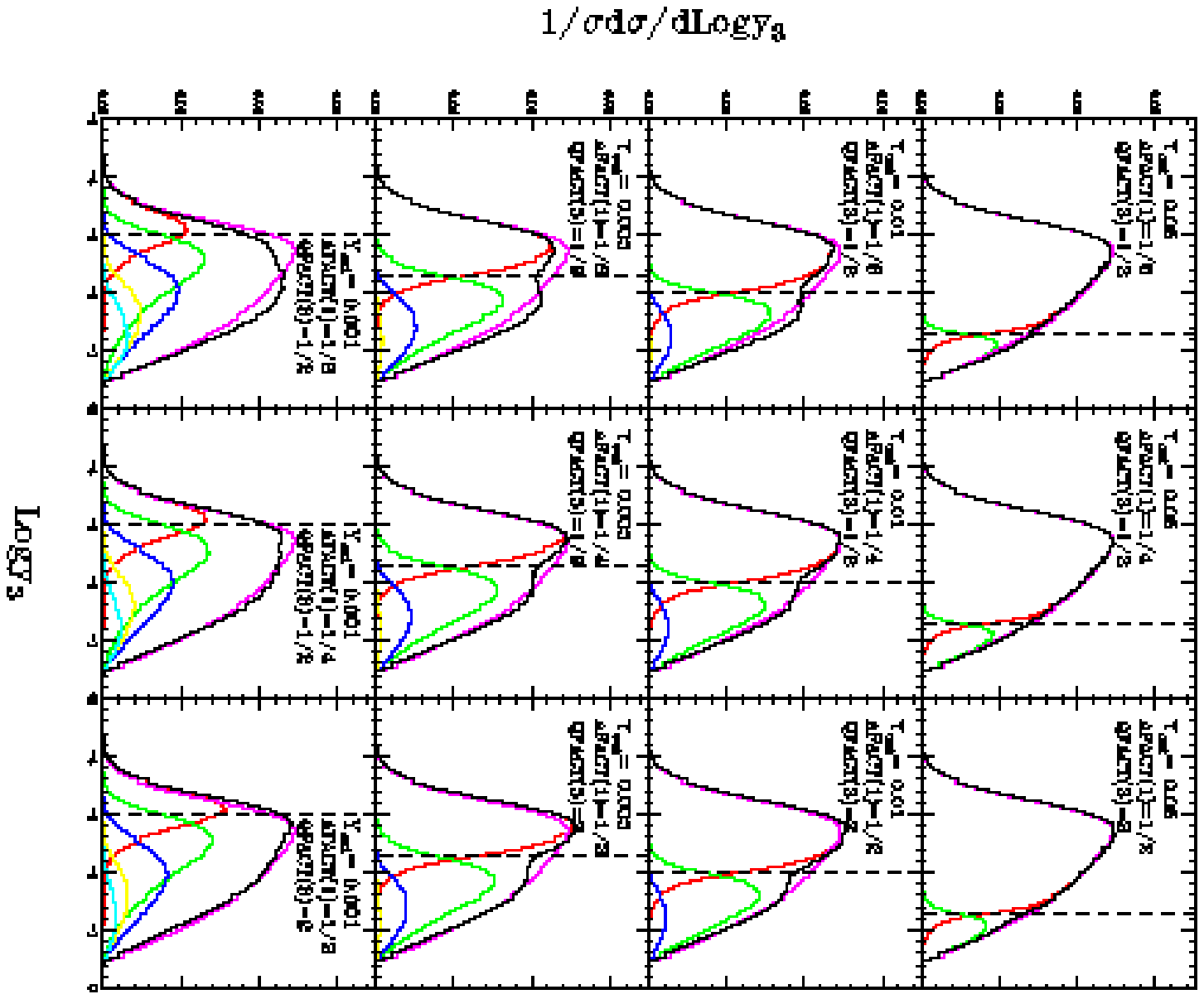}}
\Text(0,375)[rc]{\black ME}
\Text(0,350)[rc]{\magenta HW}
\Text(0,325)[rc]{\red 2 jets}
\Text(0,300)[rc]{\blue 3 jets}
\Text(0,275)[rc]{\green 4 jets}
\Text(0,250)[rc]{\yellow 5 jets}
\Text(0,225)[rc]{\cyan 6 jets}
\Text(200,425)[c]{\sf \HW-CKKW (Hadron Level)}
\end{picture}

\caption{Effect of varying the matching scale
         using NLL Sudakov form factors
         on the  hadron-level 
         differential
         cross section $\frac{1}{\sigma}\frac{d\sigma}{d\log y_3}$ in 
         $\ee$ collisions at $\sqrt{s}=M_Z$. The parameters
         were set so that {\tt QFACT(1)=QFACT(2)=1/2} and {\tt AFACT(1)=AFACT(2)}.
         The default result of \HW\ is shown as a magenta line, 
         the result of the \CK\ algorithm is shown as a black line.
         The contribution to the \CK\ result of the different jet multiplicities
         are also shown, red is the 2 jet component, green is the 3 jet component,
         blue is the 4 jet component, yellow is the 5 jet component and cyan is the
         6 jet component. The matching scale is shown as a vertical dashed line.
         The {\tt IFINAL=3} option was used for the highest multiplicity matrix
         element.}
\label{fig:eelepy3had}}

\FIGURE[!ht]{
\begin{picture}(400,400)
\put(50,0){\includegraphics[bbllx=100,bblly=0,bburx=550,bbury=550,angle=90]{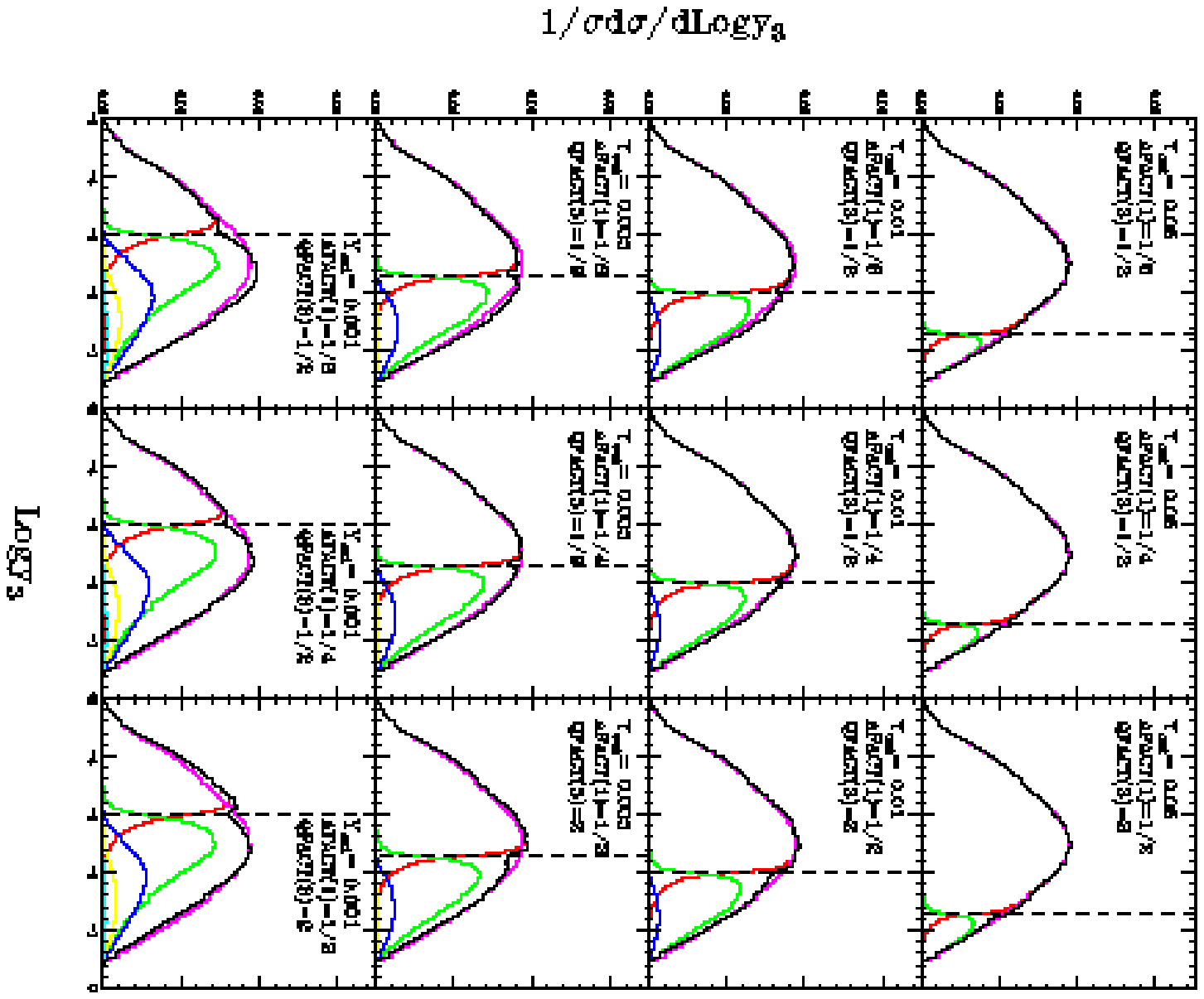}}
\Text(0,375)[rc]{\black ME}
\Text(0,350)[rc]{\magenta HW}
\Text(0,325)[rc]{\red 2 jets}
\Text(0,300)[rc]{\blue 3 jets}
\Text(0,275)[rc]{\green 4 jets}
\Text(0,250)[rc]{\yellow 5 jets}
\Text(0,225)[rc]{\cyan 6 jets}
\Text(200,425)[c]{\sf \HW-CKKW (Hadron Level)}
\end{picture}
\caption{Effect of varying the matching scale
         using NLL Sudakov form factors
         on the hadron-level differential
         cross section $\frac{1}{\sigma}\frac{d\sigma}{d\log y_3}$ in 
         $\ee$ collisions at $\sqrt{s}=500$~GeV. The parameters
         were set so that {\tt QFACT(1)=QFACT(2)=1/2} and 
         {\tt AFACT(1)=AFACT(2)}.
         The default result of \HW\ is shown as a magenta line, 
         the result of the \CK\ algorithm is shown as a black line.
         The contribution to the \CK\ result of the different jet multiplicities
         are also shown, red is the 2 jet component, 
         green is the 3 jet component,
         blue is the 4 jet component, 
         yellow is the 5 jet component and cyan is the
         6 jet component. The matching scale is shown as a 
         vertical dashed line.
         The {\tt IFINAL=3} option was used for the highest multiplicity matrix
         element.}
\label{fig:eenlcy3had}}

\FIGURE[!ht]{
\resizebox{.95\textwidth}{!}{
\includegraphics{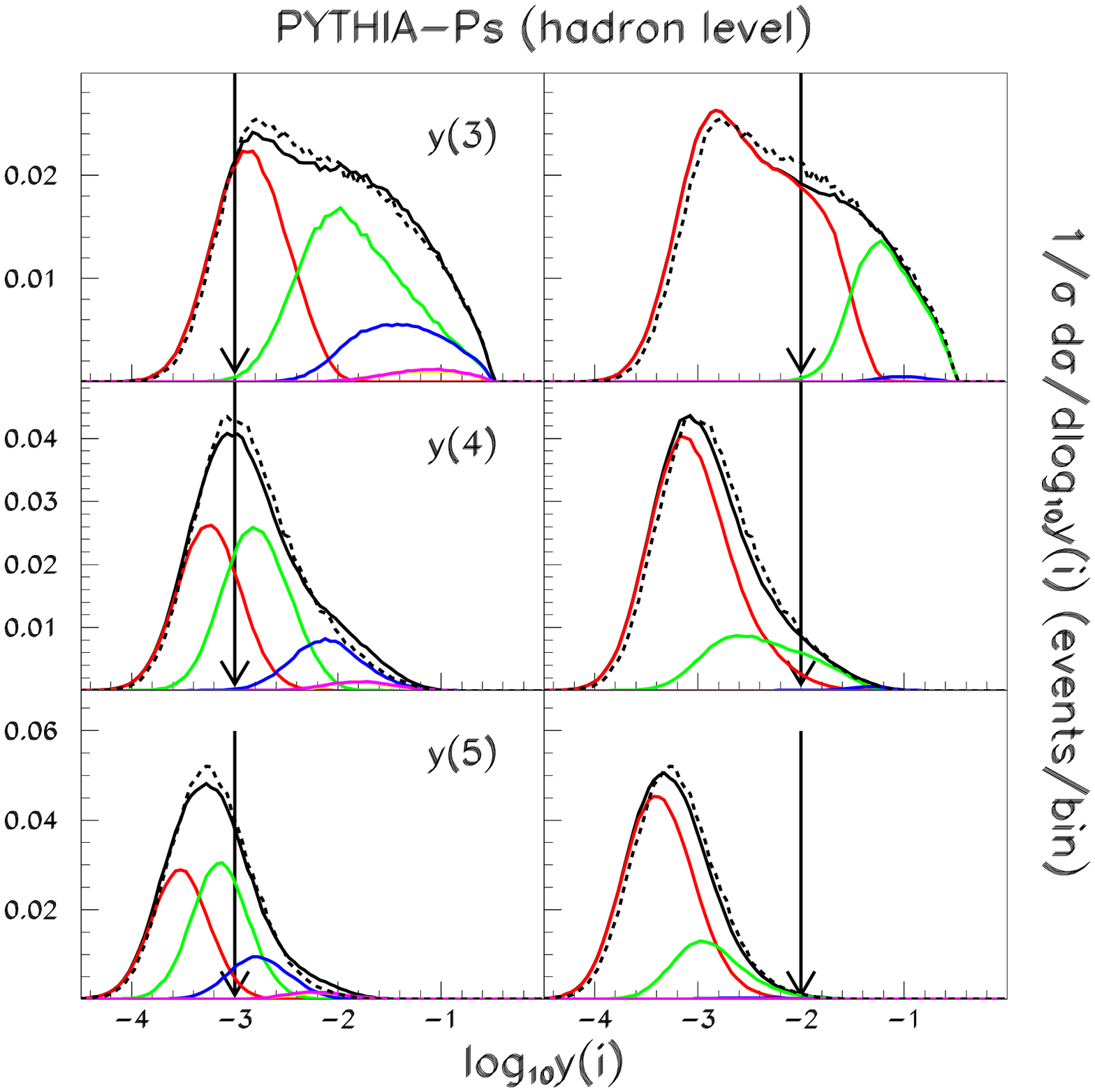}
}
\caption{Effect of varying the matching scale
         using the pseudo-shower procedure
         on the hadron-level 
         differential
         cross section $\frac{1}{\sigma}\frac{d\sigma}{d\log y_i}$ in 
         $\ee$ collisions at $\sqrt{s}=M_Z$, for $i=3,4$ and 5. 
         The default result of \PY\ is shown as a dashed line, while
         the result of the pseudo-shower algorithm is shown as a 
         solid black line.
         The contribution to the pseudo-shower result 
          from the two (red), three (green), four (blue),
         five (yellow) and six (magenta) parton
         components is also shown.
         The matching scales 
         $10^{-3} \sim (2.88)^2$ GeV$^2$ and
         $10^{-2} \sim (9.12)^2$ GeV$^2$
  are shown as vertical arrows.}
\label{fig:ycut001}
}

\FIGURE[!ht]{
\begin{picture}(400,400)
\put(0,0){\includegraphics[width=0.95\textwidth,angle=90]{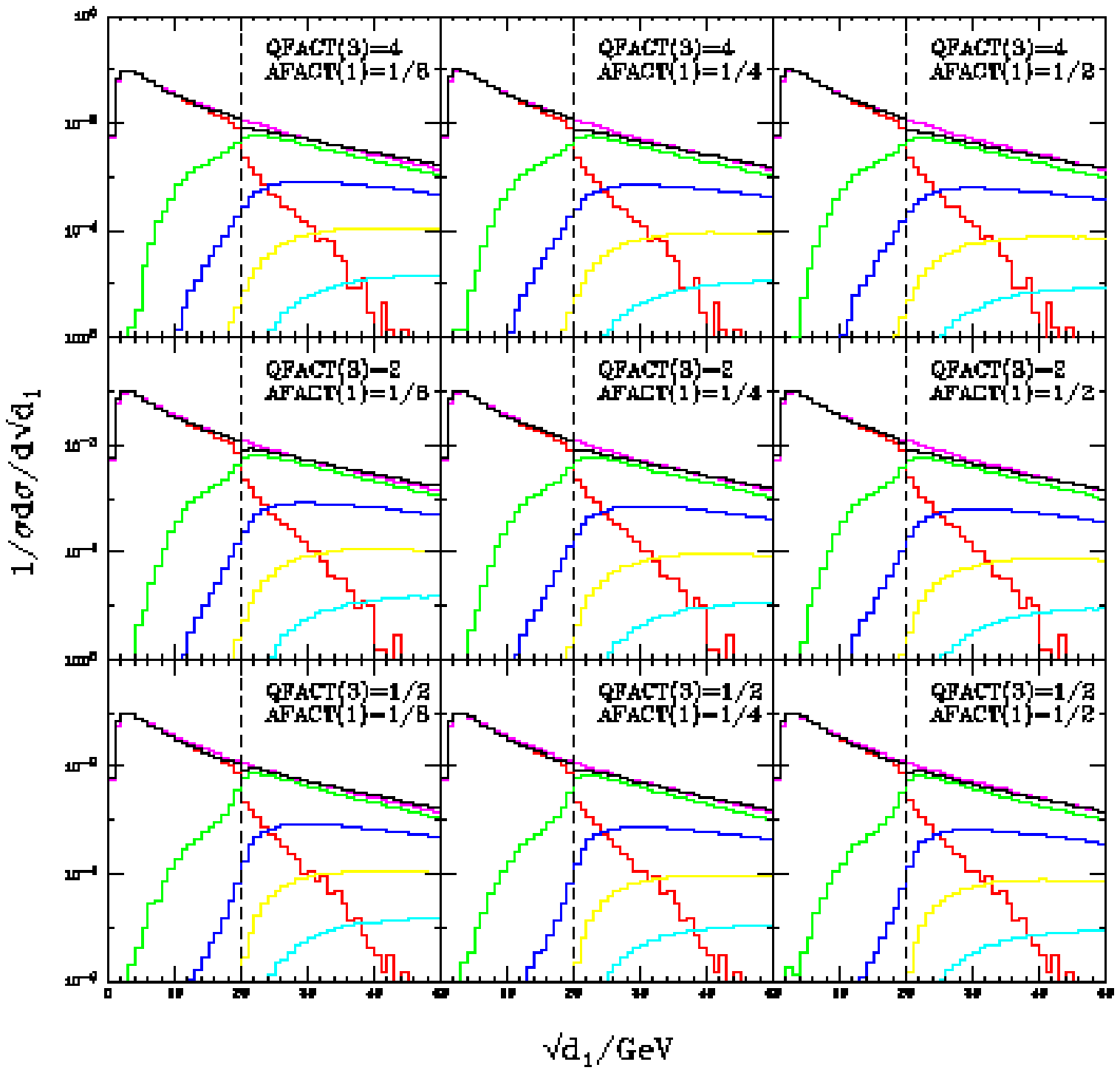}}
\Text(0,375)[rc]{\black ME}
\Text(0,350)[rc]{\magenta HW}
\Text(0,325)[rc]{\red 2 jets}
\Text(0,300)[rc]{\blue 3 jets}
\Text(0,275)[rc]{\green 4 jets}
\Text(0,250)[rc]{\yellow 5 jets}
\Text(0,225)[rc]{\cyan 6 jets}
\Text(200,425)[c]{\sf \HW-CKKW (Parton Level)}
\end{picture}

\caption{Effect of varying the scales for $\as$ and the minimum starting
         scale for the shower on the parton-level differential cross
         section $\frac{1}{\sigma}\frac{d\sigma}{d\sqrt{d_1}}$ for W production
         at the Tevatron
         for a centre-of-mass energy of $1.96$~TeV. The parameters
         were set so that {\tt QFACT(1)=QFACT(2)=1/2},
         {\tt QFACT(3)=QFACT(4)} and {\tt AFACT(1)=AFACT(2)}.
         The default result of \HW\ is shown as a magenta line, 
         the result of the \CK\ algorithm is shown as a black line.
         The contribution to the \CK\ result of the different jet multiplicities
         are also shown, red is the 0 jet component, green is the 1 jet component,
         blue is the 2 jet component, yellow is the 3 jet component and cyan is the
         4 jet component. The matching scale is shown as a vertical dashed line.
         The {\tt IFINAL=3} option was used for the highest multiplicity matrix
         element.}
\label{fig:Wd1a}}

\FIGURE[!ht]{
\begin{picture}(400,400)
\put(0,0){\includegraphics[width=0.95\textwidth,angle=90]{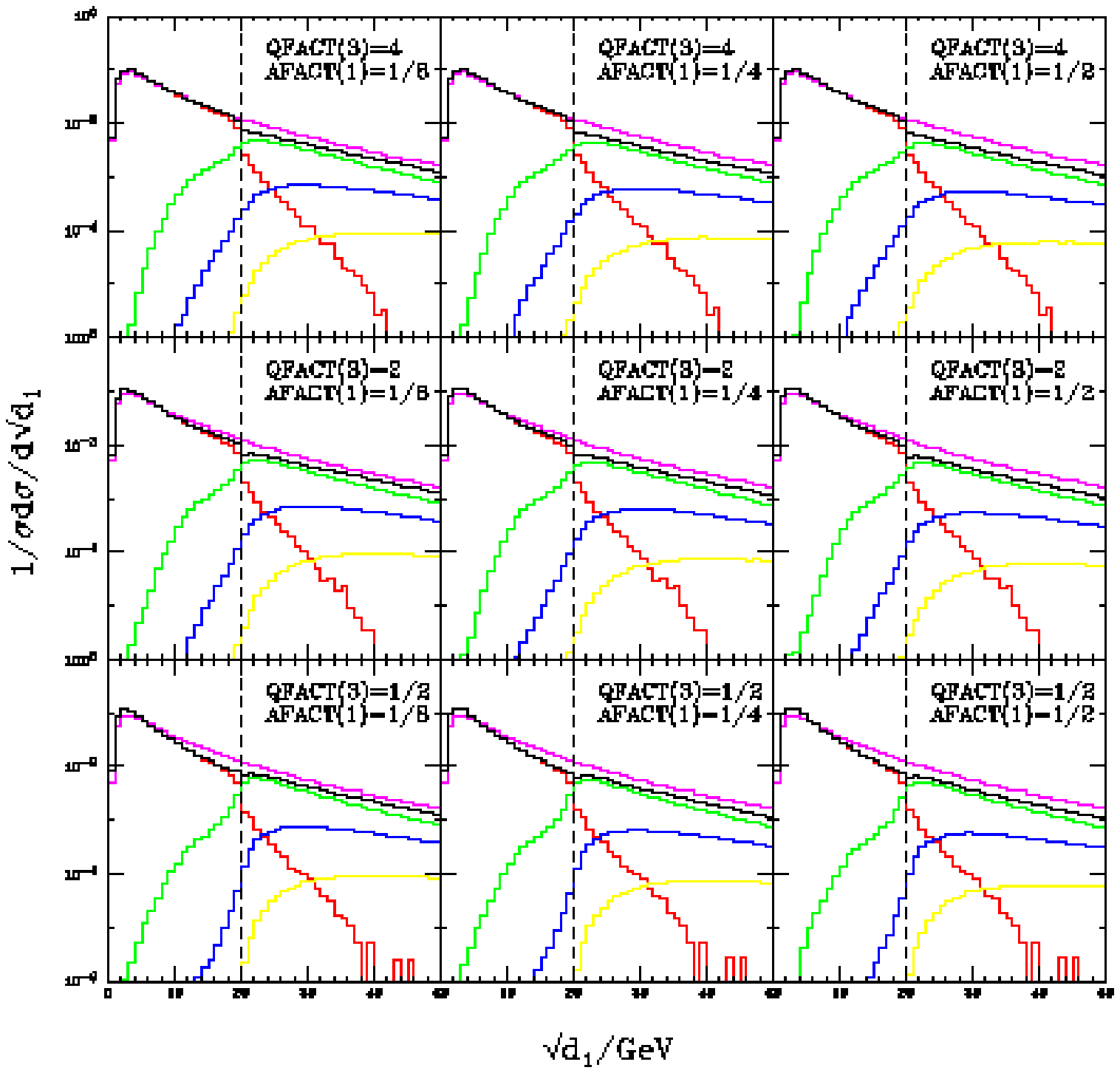}}
\Text(0,375)[rc]{\black ME}
\Text(0,350)[rc]{\magenta HW}
\Text(0,325)[rc]{\red 2 jets}
\Text(0,300)[rc]{\blue 3 jets}
\Text(0,275)[rc]{\green 4 jets}
\Text(0,250)[rc]{\yellow 5 jets}
\Text(0,225)[rc]{\cyan 6 jets}
\Text(200,425)[c]{\sf \HW-CKKW (Parton Level)}
\end{picture}

\caption{Effect of varying the scales for $\as$ and the minimum starting
         scale for the shower on the parton-level differential cross
         section $\frac{1}{\sigma}\frac{d\sigma}{d\sqrt{d_1}}$ for Z~production
         at the Tevatron
         for a centre-of-mass energy of $1.96$~TeV. The parameters
         were set so that {\tt QFACT(1)=QFACT(2)=1/2},
         {\tt QFACT(3)=QFACT(4)} and {\tt AFACT(1)=AFACT(2)}.
         The default result of \HW\ is shown as a magenta line, 
         the result of the \CK\ algorithm is shown as a black line.
         The contribution to the \CK\ result of the different jet multiplicities
         are also shown, red is the 0 jet component, green is the 1 jet component,
         blue is the 2 jet component, yellow is the 3 jet component and cyan is the
         4 jet component. The matching scale is shown as a vertical dashed line.
         The {\tt IFINAL=3} option was used for the highest multiplicity matrix
         element.}
\label{fig:Zd1a}}

\clearpage

\FIGURE[H]{
\begin{picture}(400,400)
\put(0,0){\includegraphics[width=0.95\textwidth,angle=90]{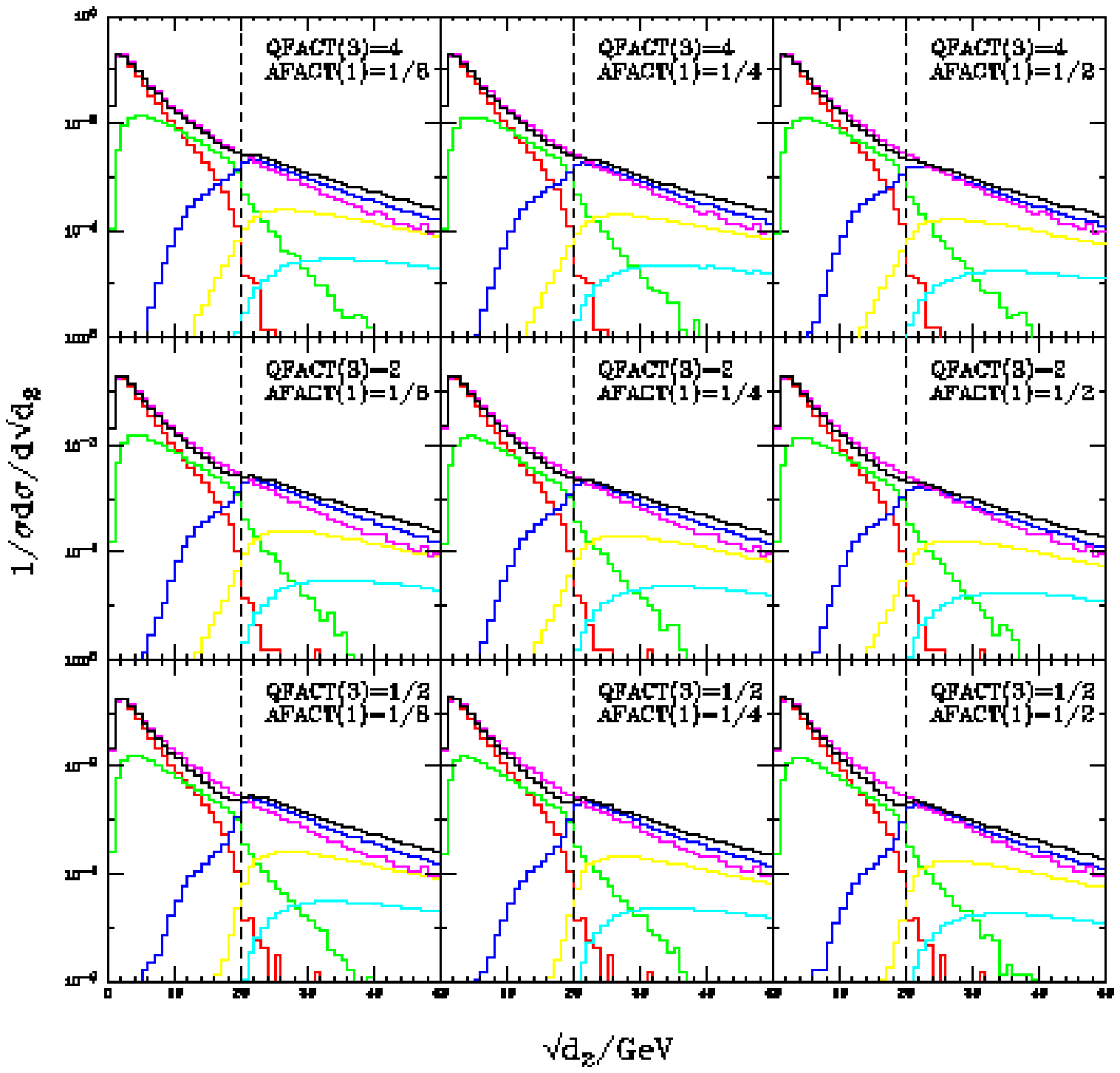}}
\Text(0,375)[rc]{\black ME}
\Text(0,350)[rc]{\magenta HW}
\Text(0,325)[rc]{\red 2 jets}
\Text(0,300)[rc]{\blue 3 jets}
\Text(0,275)[rc]{\green 4 jets}
\Text(0,250)[rc]{\yellow 5 jets}
\Text(0,225)[rc]{\cyan 6 jets}
\Text(200,425)[c]{\sf \HW-CKKW (Parton Level)}
\end{picture}

\caption{Effect of varying the scales for $\as$ and the minimum starting
         scale for the shower on the parton-level differential cross
         section $\frac{1}{\sigma}\frac{d\sigma}{d\sqrt{d_2}}$ for W production
         at the Tevatron
         for a centre-of-mass energy of $1.96$~TeV. The parameters
         were set so that {\tt QFACT(1)=QFACT(2)=1/2},
         {\tt QFACT(3)=QFACT(4)}and {\tt AFACT(1)=AFACT(2)}.
         The default result of \HW\ is shown as a magenta line, 
         the result of the \CK\ algorithm is shown as a black line.
         The contribution to the \CK\ result of the different jet multiplicities
         are also shown, red is the 0 jet component, 
         green is the 1 jet component,
         blue is the 2 jet component, 
         yellow is the 3 jet component and cyan is the
         4 jet component. 
         The matching scale is shown as a vertical dashed line.
         The {\tt IFINAL=3} option was used for the highest multiplicity matrix
         element.}
\label{fig:Wd2a}
}

\FIGURE[!ht]{
\includegraphics[width=.95\textwidth,angle=0]{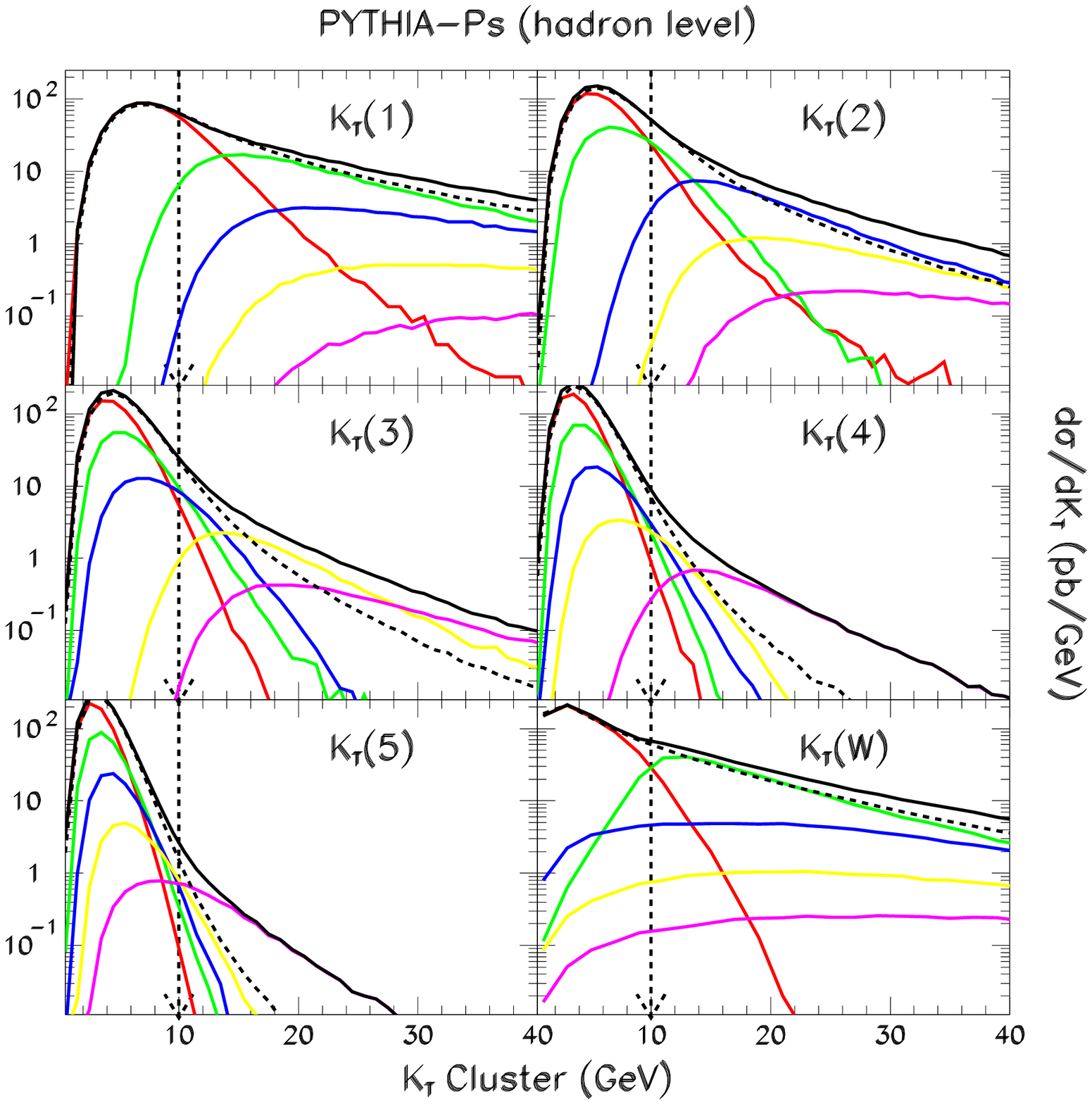}
\caption{
Differential $k_{Ti}$-cluster distributions 
  $d\sigma/d{k_{Ti}}$ 
         at the hadron level generated
         with the pseudo-shower procedure for
         $\rm{p}\bar{\rm{p}}\to W^{+}+X$ collisions at $\sqrt{s}=1.96$ TeV, 
        for $i=1-5$ and also showing the W$^+$ boson transverse momentum.
         The default result of \PY\ is shown as a dashed line, while
         the result of the pseudo-shower algorithm is shown as a 
         solid black line.
         The contribution to the pseudo-shower result 
          from the two (red), three (green), four (blue),
         five (yellow) and six (magenta) parton
         components is also shown.
         The matching scale 10 GeV is
         shown as a vertical arrow.}
\label{fig:W10sm}
}

\FIGURE[!ht]{
\includegraphics[width=.95\textwidth]{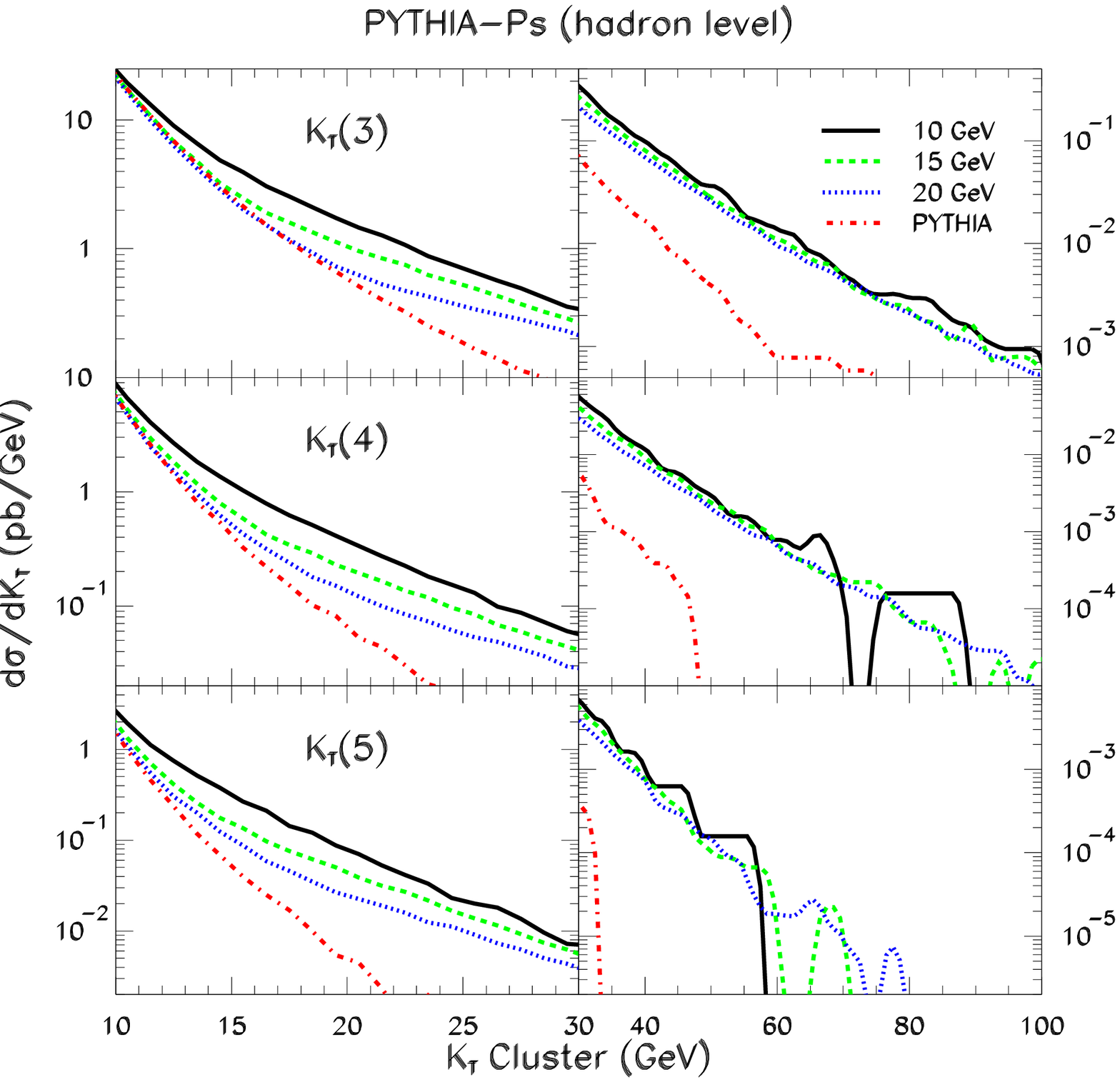}
\caption{
Differential $k_{Ti}$-cluster distributions 
  $d\sigma/d{k_{Ti}}$ for $i=3,4$ and 5
         at the hadron level generated
         with the pseudo-shower procedure for
         $\rm{p}\bar{\rm{p}}\to W^{+}+X$ collisions at $\sqrt{s}=1.96$ TeV. 
         The default result of \PY\ is shown as the
         (red) dash-dot line.  The dependence on different
         matching scales is shown:
         10 GeV (black, solid); 15 GeV (green, dash); and
         20 GeV (blue, short dash).
         The highest multiplicity matrix element used
         in each case is W$^{+}+4$ partons.}
\label{fig:W345sm}
}

\FIGURE[!ht]{
\includegraphics[width=.95\textwidth]{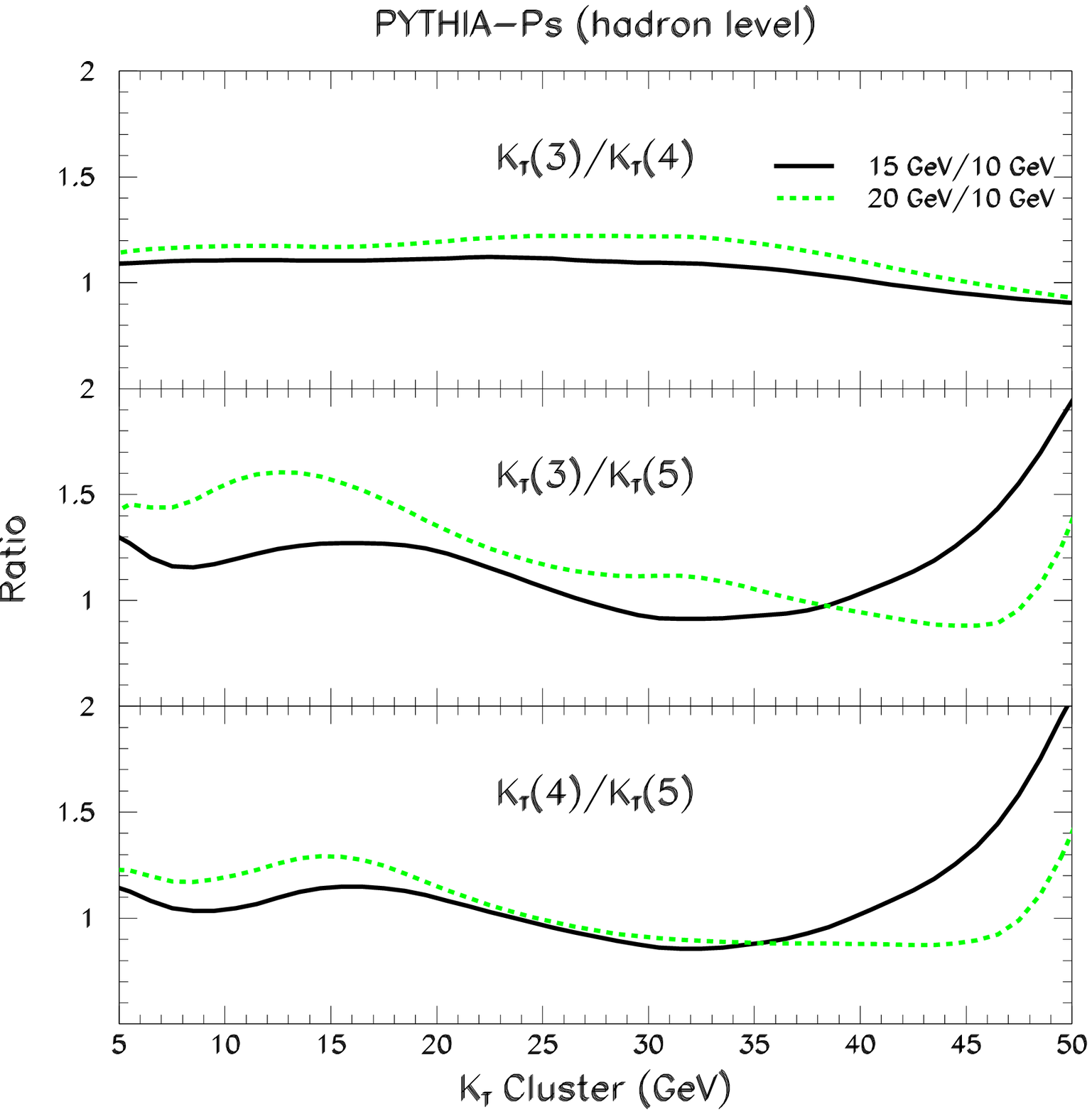}
\caption{Comparison of the ratio of $k_T$ cluster
distributions in Fig.~\ref{fig:W345sm} for 
the same matching scales.}
\label{fig:ratio}
}

\FIGURE[!ht]{
\includegraphics[width=.95\textwidth]{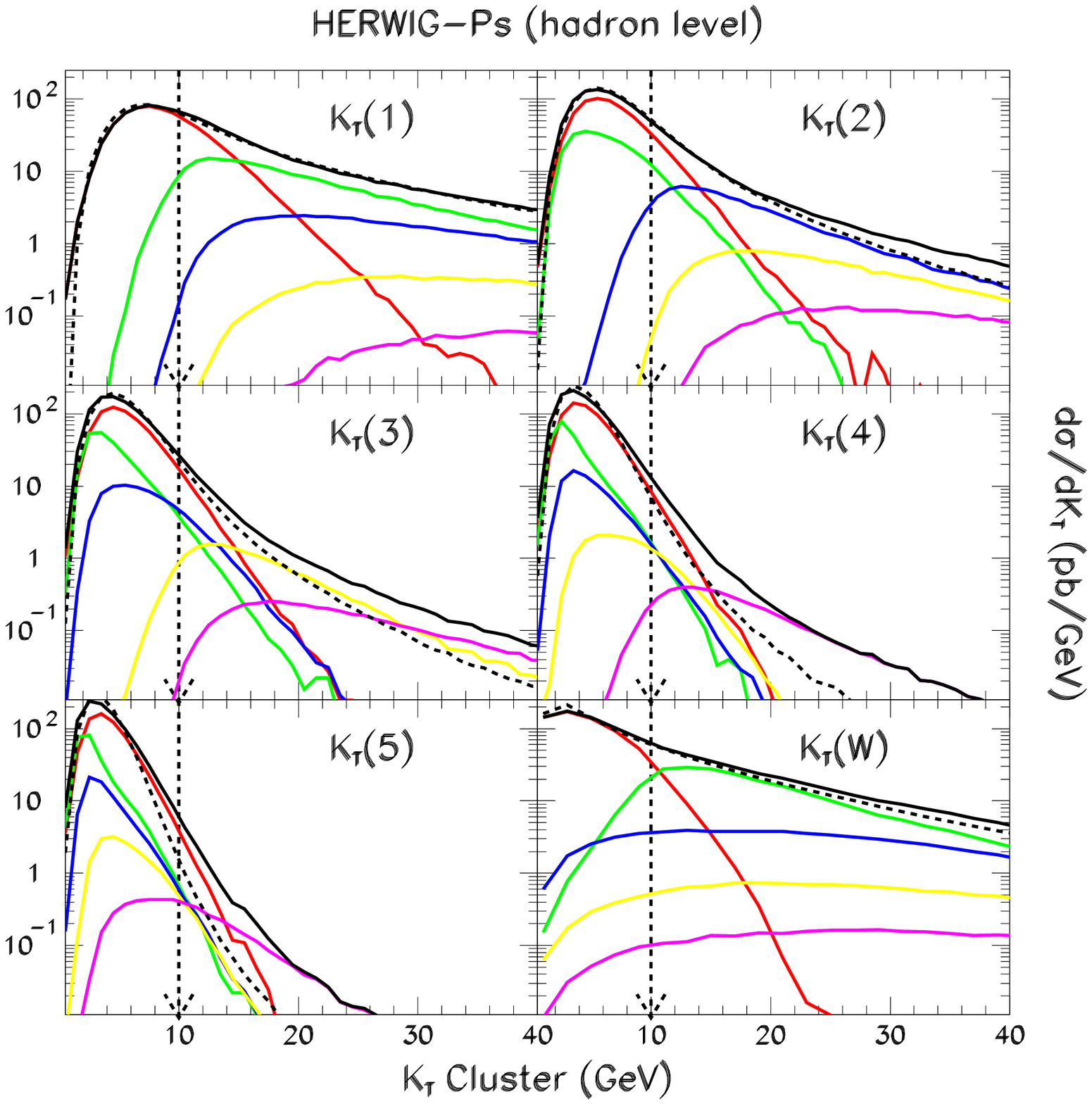}
\caption{Same as Fig.~\ref{fig:W10sm}, but using
\HW\ in the pseudo-shower procedure.}
\label{fig:W10sm_hw}
}

\FIGURE[!ht]{
\includegraphics[width=.95\textwidth]{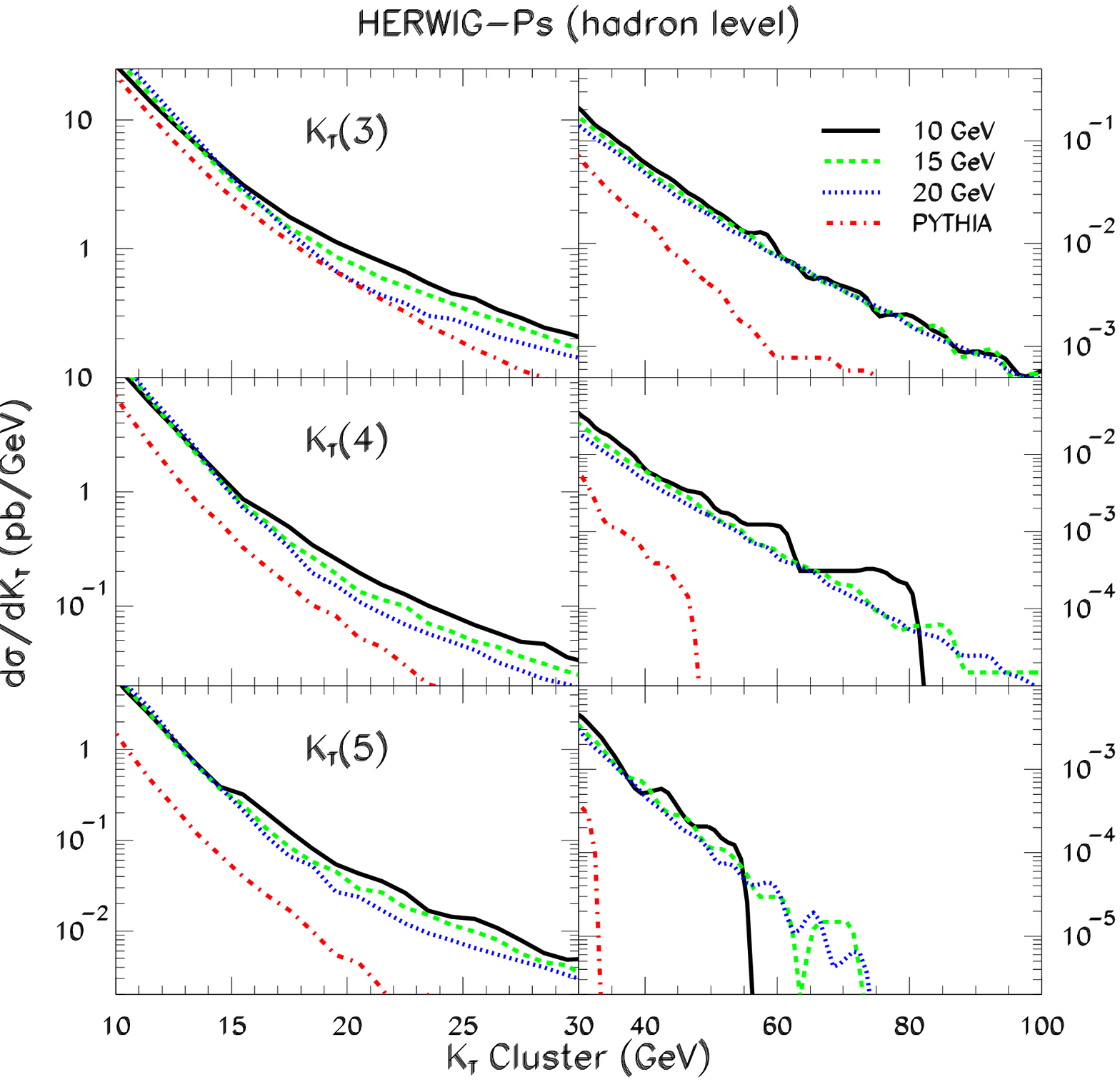}
\caption{Same as Fig.~\ref{fig:W345sm}, but using
\HW\ in the pseudo-shower procedure.}
\label{fig:W345sm_hw}
}

\FIGURE[!ht]{
\includegraphics[width=.95\textwidth]{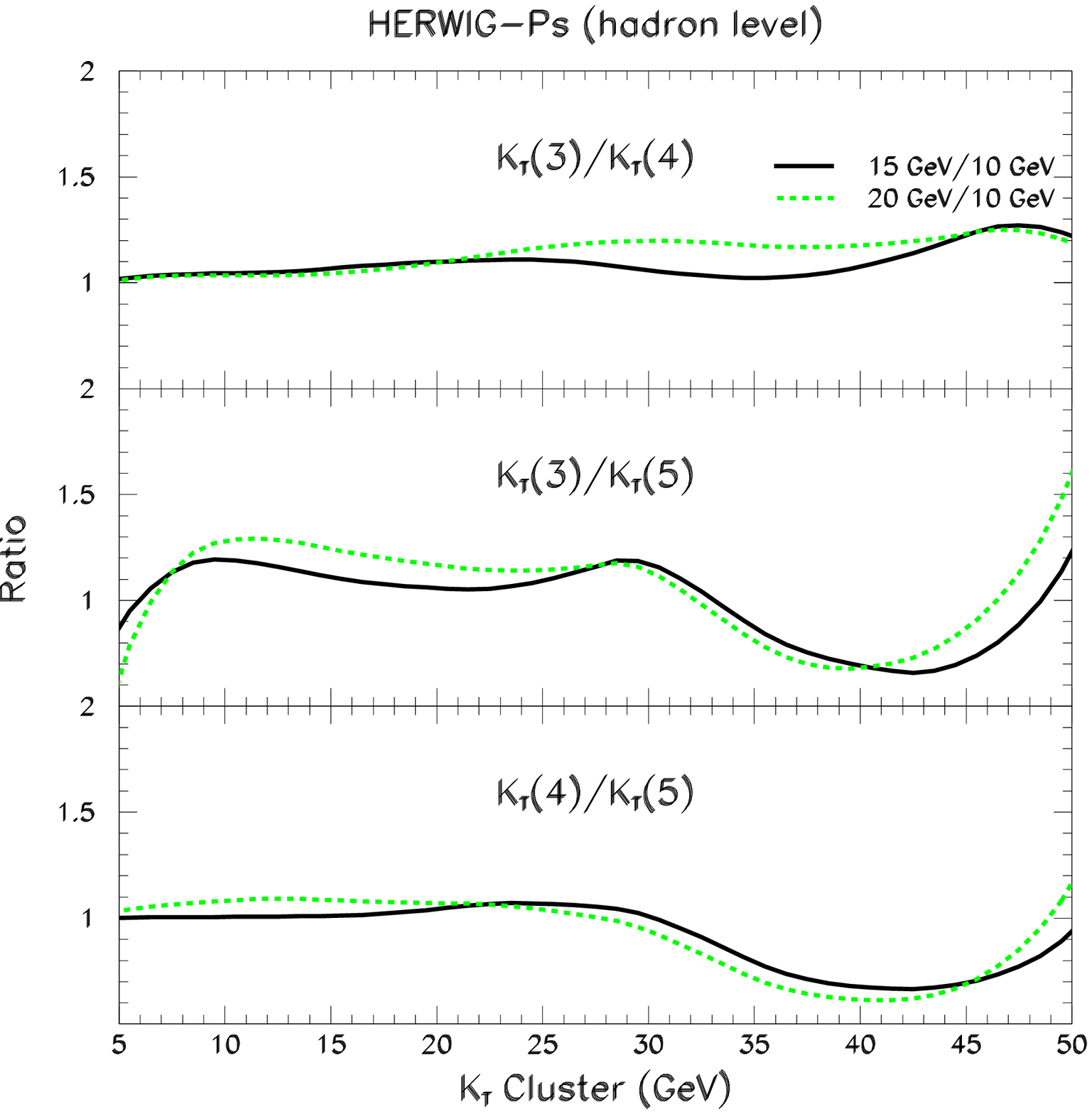}
\caption{Same as Fig.~\ref{fig:ratio}, but using
\HW\ in the pseudo-shower procedure.}
\label{fig:ratio_hw}
}

\FIGURE[!ht]{
\includegraphics[width=.95\textwidth]{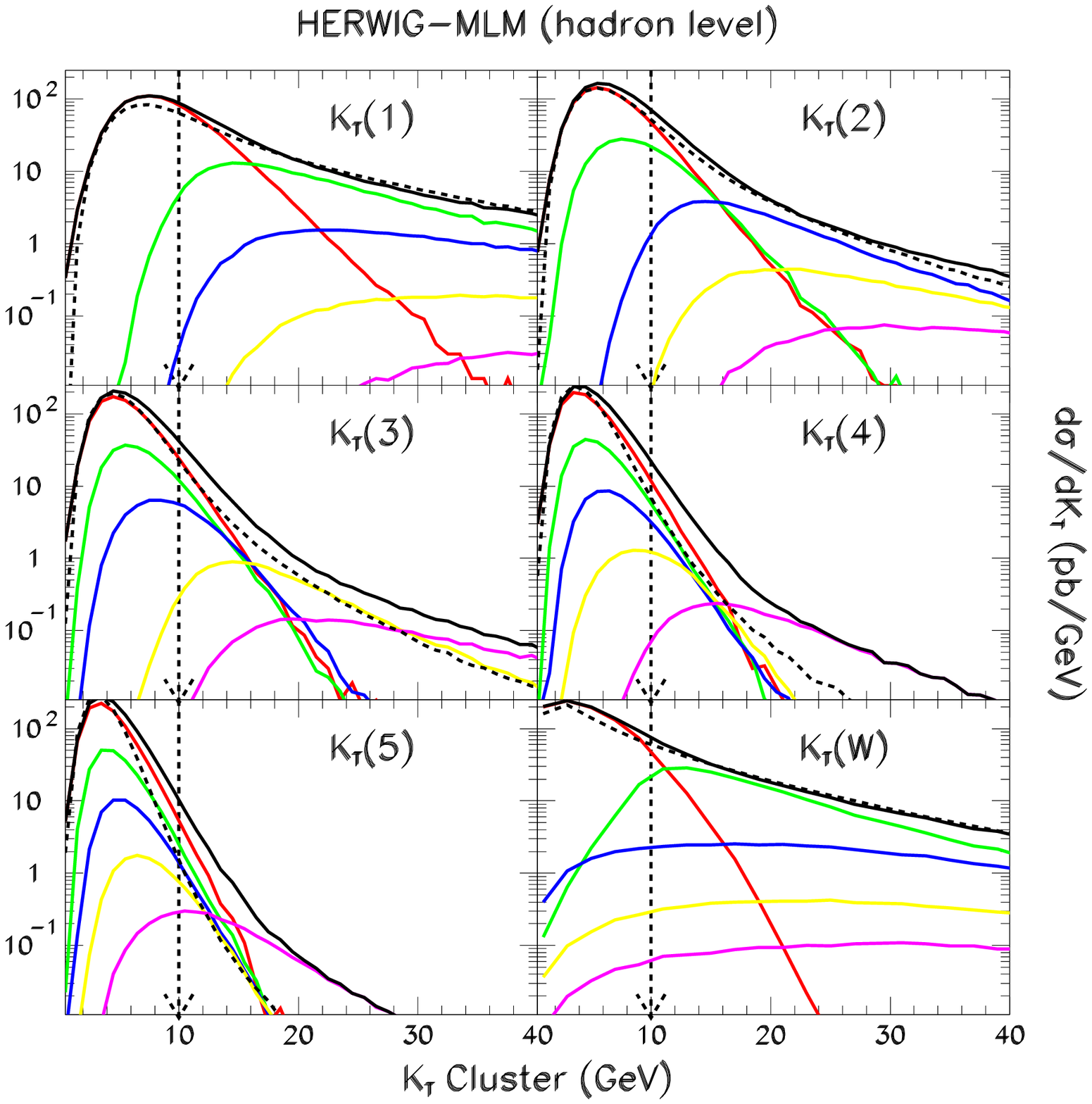}
\caption{Same as Fig.~\ref{fig:W10sm}, but using
\HW\ in the MLM procedure.}
\label{fig:mlm_kt1}
}

\FIGURE[!ht]{
\includegraphics[width=.95\textwidth]{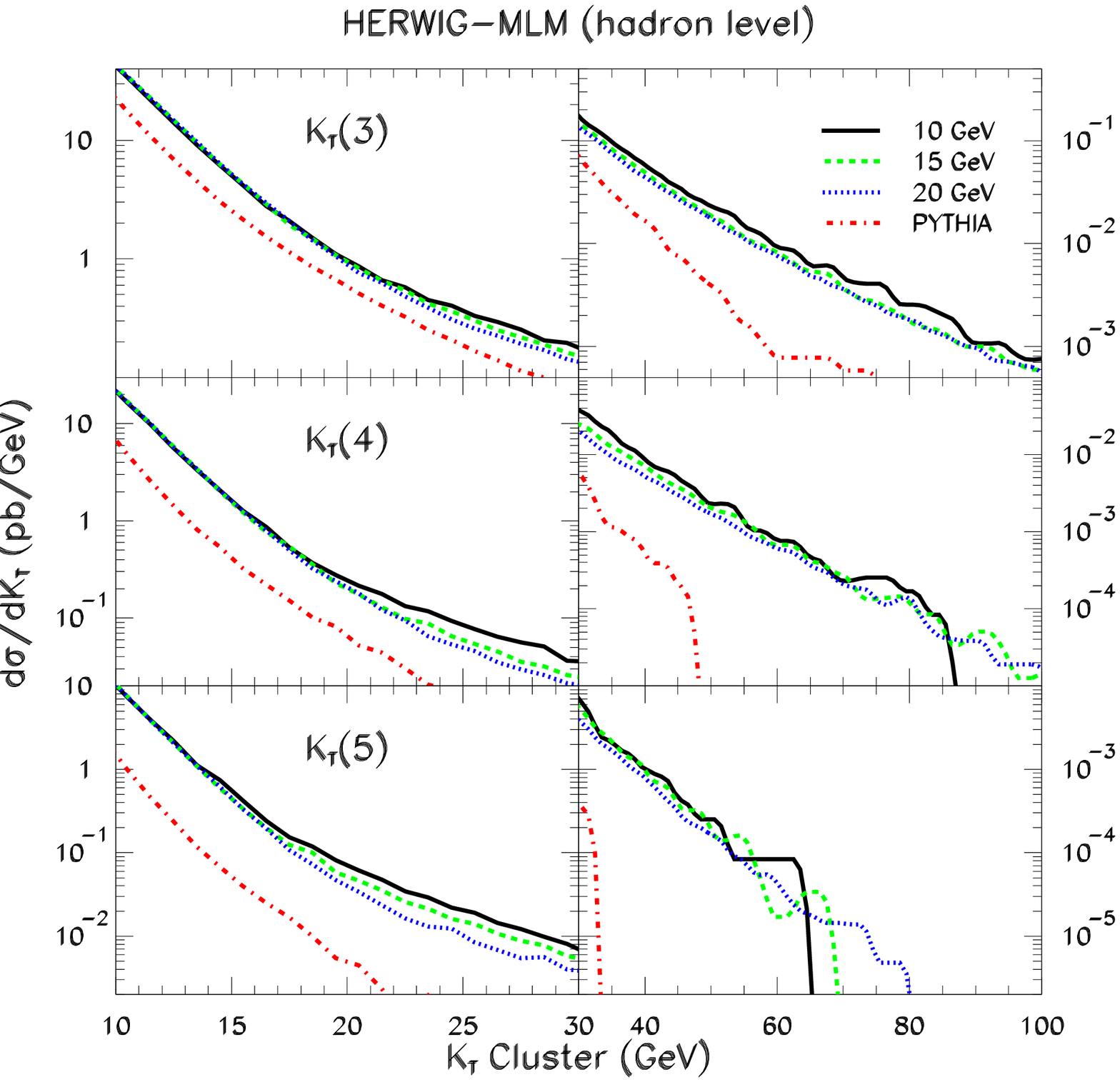}
\caption{Same as Fig.~\ref{fig:W345sm}, but using
\HW\ in the MLM procedure.}
\label{fig:mlm_kt345}
}

\FIGURE[!ht]{
\includegraphics[width=.95\textwidth]{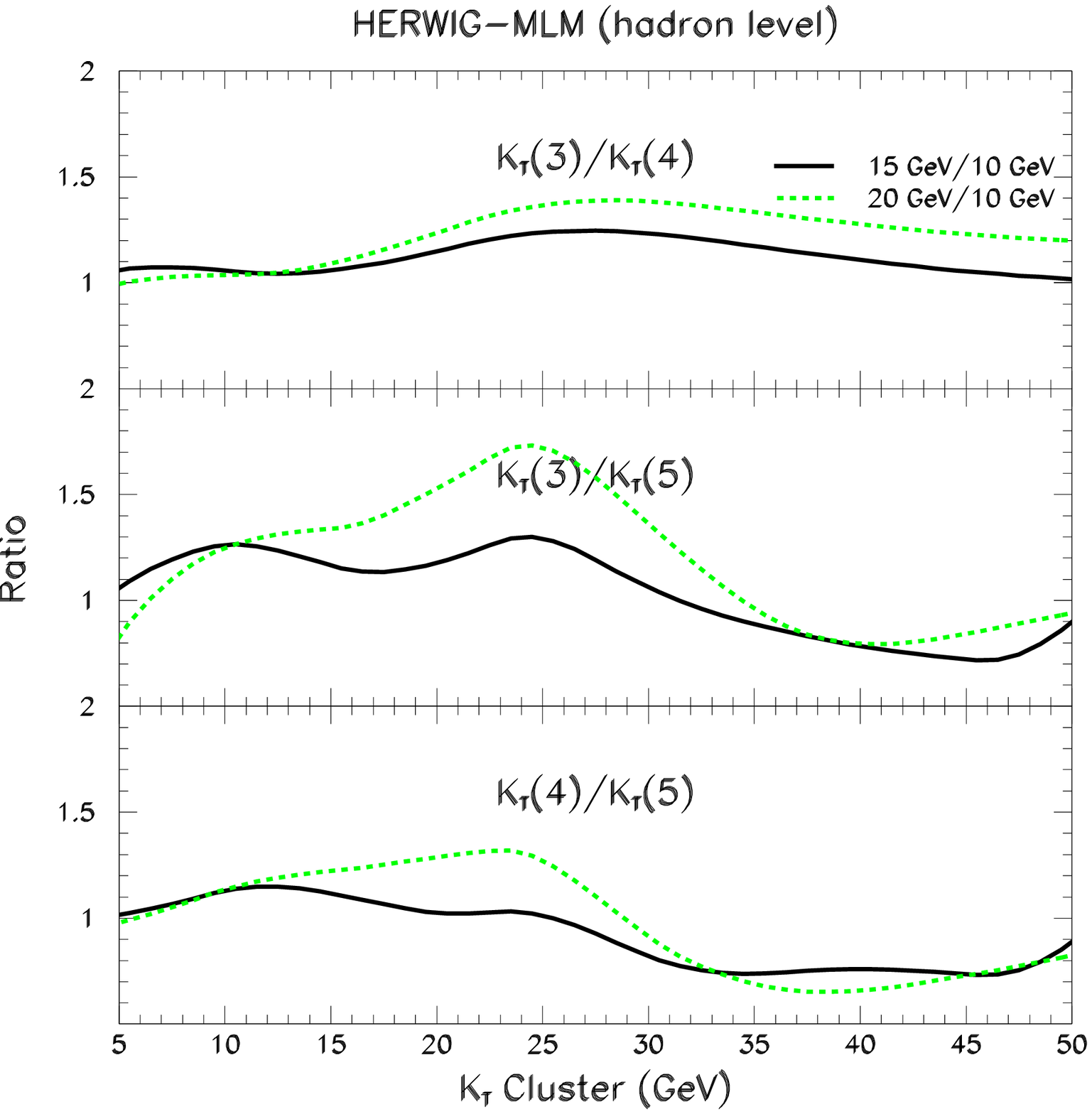}
\caption{Same as Fig.~\ref{fig:W345sm}, but using
\HW\ in the MLM procedure.}
\label{fig:mlm_ktr}
}

\FIGURE[!ht]{
\includegraphics[width=.95\textwidth]{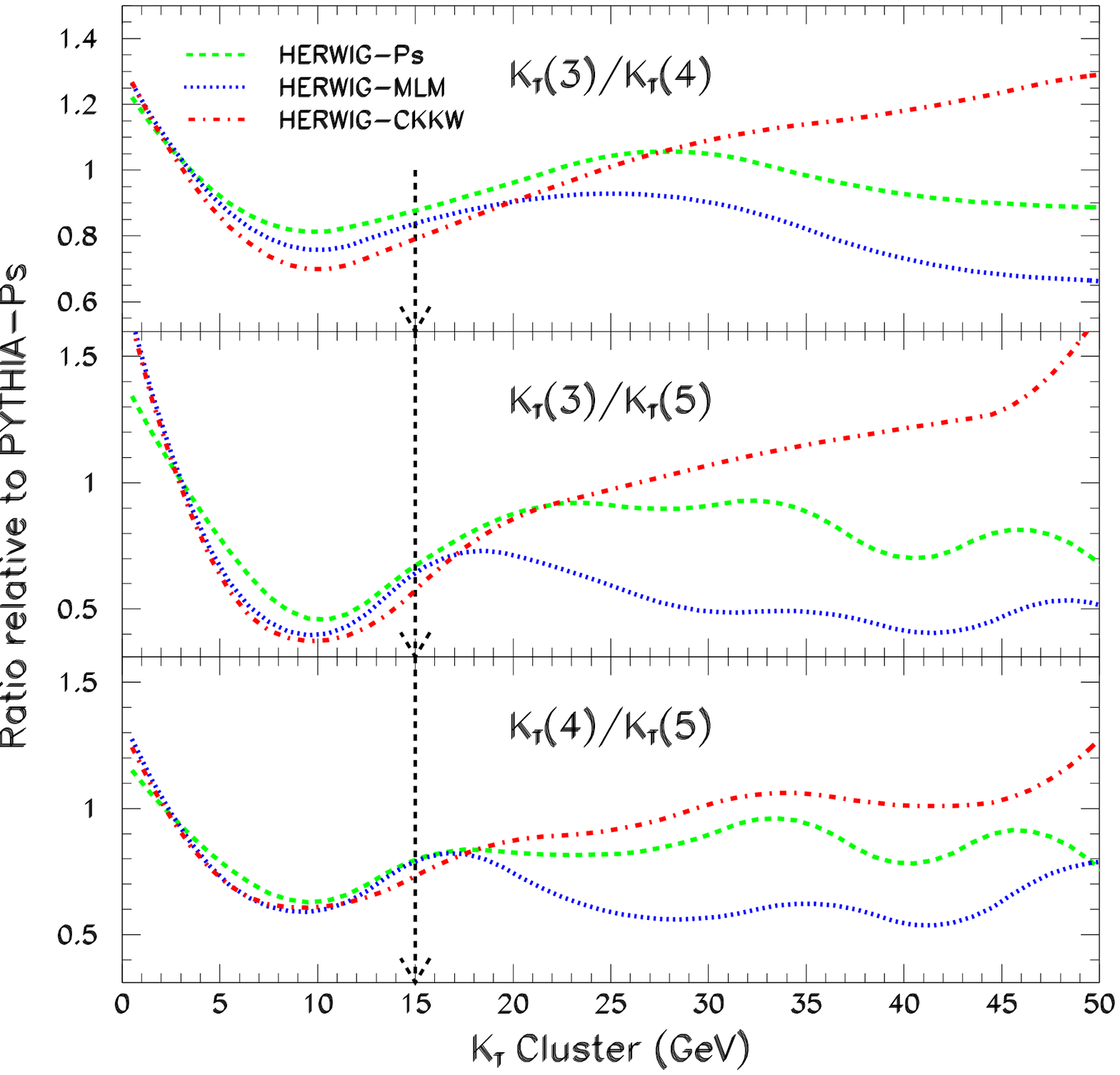}
\caption{
Similar to Fig.~\ref{fig:ratio}
but comparing the
distributions from \HW\ and PY\ using the
pseudo-shower procedure, \HW\
 using the MLM procedure, and
\HW\ using the CKKW procedure
for a matching scale of 15 GeV.}
\label{fig:compare_ratio}
}

\end{document}